\documentclass[prl,preprint,superscriptaddress,showpacs,showkeys,amsmath,amssymb,aps]{revtex4-1}

\setcitestyle{super}
\usepackage{graphicx}
\usepackage{amsmath}
\usepackage[version=3]{mhchem}
\usepackage[usenames]{color}
\usepackage{color}
\usepackage{physics}
\usepackage[colorlinks=true, linkcolor=blue, citecolor=blue, urlcolor=blue,pdftex]{hyperref}
\definecolor{newcolor}{rgb}{0.9,0,0.1}

\usepackage{orcidlink}
\usepackage{multirow}
\usepackage{siunitx}
\usepackage{float}

\renewcommand{\baselinestretch}{1.4}
\newcommand{\vthzpk}{$V_\text{THz}^\text{pk}$}

\newcommand{\qlw}{$Q_\text{LW}$}
\newcommand{\vdc}{$V_\text{dc}$}
\newcommand{\idc}{$I_\text{dc}$}
\newcommand{\didv}{d$I$/d$V$}
\newcommand{\vacse}{Vac\textsubscript{Se}}
\newcommand{\wse}{WSe\textsubscript{2}}

\newcommand{\angstrom}{Å}

\usepackage{xr}
\makeatletter
\newcommand*{\addFileDependency}[1]{% argument=file name and extension
\typeout{(#1)}% latexmk will find this if $recorder=0
\@addtofilelist{#1}
\IfFileExists{#1}{}{\typeout{No file #1.}}

\makeatother
}

\begin{document}

\title{Ultrafast Coulomb blockade in an atomic-scale quantum dot}

\author{Jonas Allerbeck\,\orcidlink{0000-0002-3912-3265} \textsuperscript{\textdagger}}
\affiliation{nanotech@surfaces Laboratory, Empa -- Swiss Federal Laboratories for Materials Science and Technology, D\"ubendorf 8600, Switzerland}

\author{Laric Bobzien\,\orcidlink{0009-0002-9792-3567} \textsuperscript{\textdagger}}
\affiliation{nanotech@surfaces Laboratory, Empa -- Swiss Federal Laboratories for Materials Science and Technology, D\"ubendorf 8600, Switzerland}

\author{Nils Krane\, \orcidlink{0000-0000-0000-0000}}
\affiliation{nanotech@surfaces Laboratory, Empa -- Swiss Federal Laboratories for Materials Science and Technology, D\"ubendorf 8600, Switzerland}

\author{S. Eve Ammerman\, \orcidlink{0000-0003-3588-2245}}
\affiliation{nanotech@surfaces Laboratory, Empa -- Swiss Federal Laboratories for Materials Science and Technology, D\"ubendorf 8600, Switzerland}

\author{Daniel E. Cintron Figueroa}
\affiliation{Department of Chemistry, The Pennsylvania State University, University Park, 16802, PA, USA}

\author{Chengye Dong}
\affiliation{Two-Dimensional Crystal Consortium and Materials Research Institute, The Pennsylvania State University, University Park, 16802, PA, USA}

\author{Joshua A. Robinson\,\orcidlink{0000-0002-1513-7187}}
\affiliation{Department of Chemistry, The Pennsylvania State University, University Park, 16802, PA, USA}
\affiliation{Two-Dimensional Crystal Consortium and Materials Research Institute, The Pennsylvania State University, University Park, 16802, PA, USA}
\affiliation{Department of Physics and Department of Materials Science and Engineering, The Pennsylvania State University, University Park, 16802, PA, USA}

\author{Bruno Schuler\,\orcidlink{0000-0002-9641-0340}}
\email[]{bruno.schuler@empa.ch}
\affiliation{nanotech@surfaces Laboratory, Empa -- Swiss Federal Laboratories for Materials Science and Technology, D\"ubendorf 8600, Switzerland}

\keywords{Charge-state lifetime, Quantum dot, Ultrafast transport, THz-STM, Coulomb blockade, Franck--Condon blockade, Transition metal dichalcogenide}

%%%%%%%%%%%%%%%%%%%%%
% ABSTRACT
%%%%%%%%%%%%%%%%%%%%%
\begin{abstract}
Controlling electron dynamics at optical clock rates is a fundamental challenge in lightwave-driven nanoelectronics. Here, we demonstrate ultrafast charge-state manipulation of individual selenium vacancies in monolayer and bilayer tungsten diselenide (\wse{}) using picosecond terahertz (THz) source pulses, focused onto the picocavity of a scanning tunneling microscope (STM). 
Using THz pump--THz probe time-domain sampling of the defect charge population, we capture atomic-scale snapshots of the transient Coulomb blockade, a signature of charge transport via quantized defect states.
We identify back tunneling of localized charges to the tip electrode as a key challenge for lightwave-driven STM when probing electronic states with charge-state lifetimes exceeding the pulse duration.
However, we show that back tunneling can be mitigated by the Franck--Condon blockade, which limits accessible vibronic transitions and promotes unidirectional charge transport.
Our rate equation model accurately reproduces the time-dependent tunneling process across the different coupling regimes.
This work builds on recent progress in imaging coherent lattice and quasiparticle dynamics with lightwave-driven STM and opens new avenues for exploring ultrafast charge dynamics in low-dimensional materials, advancing the development of lightwave-driven nanoscale electronics.
\end{abstract}

\keywords{Charge-state lifetime, Quantum dot, Ultrafast transport, THz-STM, Coulomb blockade, Franck--Condon blockade, Transition metal dichalcogenide}

\date{\today}
\pacs{}
\maketitle

%%%%%%%%%%%%%
% MAIN TEXT
%%%%%%%%%%%%%

\section{Introduction}\label{sec1}
Artificial atom qubits, such as quantum dots (QDs) and color centers, have attracted significant interest in light of their applications in quantum sensing, communication, and photonic quantum computing~\cite{Weber:2010nrz, Awschalom:2013oky, wolfowiczQuantumguidelinessolidstate2021}. In particular, the inherently strong confinement for solid-state atomic-scale defects imposes a large characteristic energy scale that leads to exceptionally long coherence times, and operability beyond cryogenic environments~\cite{maurerRoomtemperaturequantumbit2012}. Spin-selective optical decay pathways enable high-fidelity spin initialization and optical readout~\cite{doldeRoomtemperatureentanglementsingle2013}.
While optical access is advantageous for addressing individual QDs at THz to PHz frequencies~\cite{heidePetahertz2024}, electrical interfaces offer crucial benefits for scalable on-chip integration.
The primary engineering challenge lies in developing suitable leads for atomic interfaces. Although extreme ultraviolet lithography is making progress towards atomic contacts, scanning probe hydrogen resist lithography~\cite{schofieldAtomicallyPrecisePlacement2003, fuechslesingleatomtransistor2012} has already demonstrated proof-of-concept devices. 
Promising candidates for well-defined atomic interfaces are two-dimensional (2D) materials that are chemically versatile, tunable via external fields, and they can host atomic-scale QDs~\cite{liu2Dmaterialsquantum2019}. Several spin-bearing, optically active defect emitters have been identified in transition metal dichalcogenides (TMDs)~\cite{kleinSiteselectivelygeneratedphoton2019, leeSpindefectqubitstwodimensional2022} and hexagonal boron nitride (hBN)~\cite{tranQuantumemissionhexagonal2016, mendelsonIdentifyingcarbonsource2021}, including substitutional defects and vacancies. However, achieving device-scale electron transport through single defects in 2D materials has so far remained elusive.\\ 

The tunneling contact in scanning probe microscopy offers a means of studying quantized electron transport through single defects and exploring the Coulomb blockade regime at the atomic scale. Steady-state Coulomb blockade of adatoms, molecules, and defects has previously been observed in scanning tunneling microscopy (STM)~\cite{senkpielDynamicalCoulombBlockade2020, kaiserChargestate2023, bobzienLayerDependent2024}. Dangling bonds of a H-terminated Si surface have been demonstrated to exhibit quantized transport~\cite{fuechslesingleatomtransistor2012,taucerSingleElectron2014} and Coulomb blockade~\cite{fuechslesingleatomtransistor2012}. Similarly, chalcogen vacancy defects in TMDs introduce discrete defect states deep in the band gap of the host material, with narrow defect resonances followed by vibronic sidebands~\cite{schulerLargeSpinOrbitSplitting2019,cochraneSpindependent2021}. These defect states can be probed individually via scanning tunneling spectroscopy (STS), revealing the influence of spin-orbit coupling~\cite{schulerLargeSpinOrbitSplitting2019} and layer-dependent binding energy~\cite{bobzienLayerDependent2024,stolzLayerdependentSchottkycontact2022}. 
To operate and control single-electron dynamics in these atomic QDs at their natural time scale, ultrafast probes are required. Recent technological advancements in ultrafast probes now bring this vision within reach, making it possible to capture 'movies' with simultaneous picosecond temporal resolution, millielectronvolt energy resolution, and picometer spatial precision~\cite{cockerNanoscaleterahertzscanning2021, boschiniTimeresolvedARPESstudies2024}. 
Ultrafast STM has resolved charge carrier dynamics in semiconductors~\cite{cockerultrafastterahertzscanning2013,jelicUltrafast2017}, coherent lattice dynamics of bulk crystals~\cite{liuNanoscalecoherentphonon2022,jelicTerahertz2024}, single molecules~\cite{cockerTrackingultrafastmotion2016}, and atomic defects~\cite{roelckeUltrafast2024, jelicAtomicscale2024}. However, the direct observation of ultrafast electron dynamics at the atomic scale has remained an open challenge.\\

Here, we address this challenge by demonstrating ultrafast Coulomb blockade at an atomic defect in a two-dimensional semiconductor by lightwave-driven STM (LW-STM). By applying tailored single-cycle THz pulses to the STM tip, we achieve a controlled population of atomic defect orbitals and trace their subsequent depletion in the time domain. The temporal evolution of the Coulomb blockade quantifies the charge-state lifetime of selenium vacancies (\vacse{}) in WSe$_2$/graphene heterostructures. By controlling the coupling strength of the tunnel junction via dc bias and tip--sample distance, along with a theoretical model utilizing the master equation, we identify the Franck--Condon blockade as a key mechanism to prevent back tunneling to the source electrode. 
This study represents the first direct observation of ultrafast electron dynamics at the atomic scale with LW-STM, opening new avenues for advancing our microscopic understanding and control of ultrafast electron transport at atomic length scales.

\section{Coulomb blockade at a singe Selenium vacancy}\label{sec2}

%%%%%%% FIGURE 1 %%%%%%%%
\begin{figure}[t]
\centering
\includegraphics[width=1\textwidth]{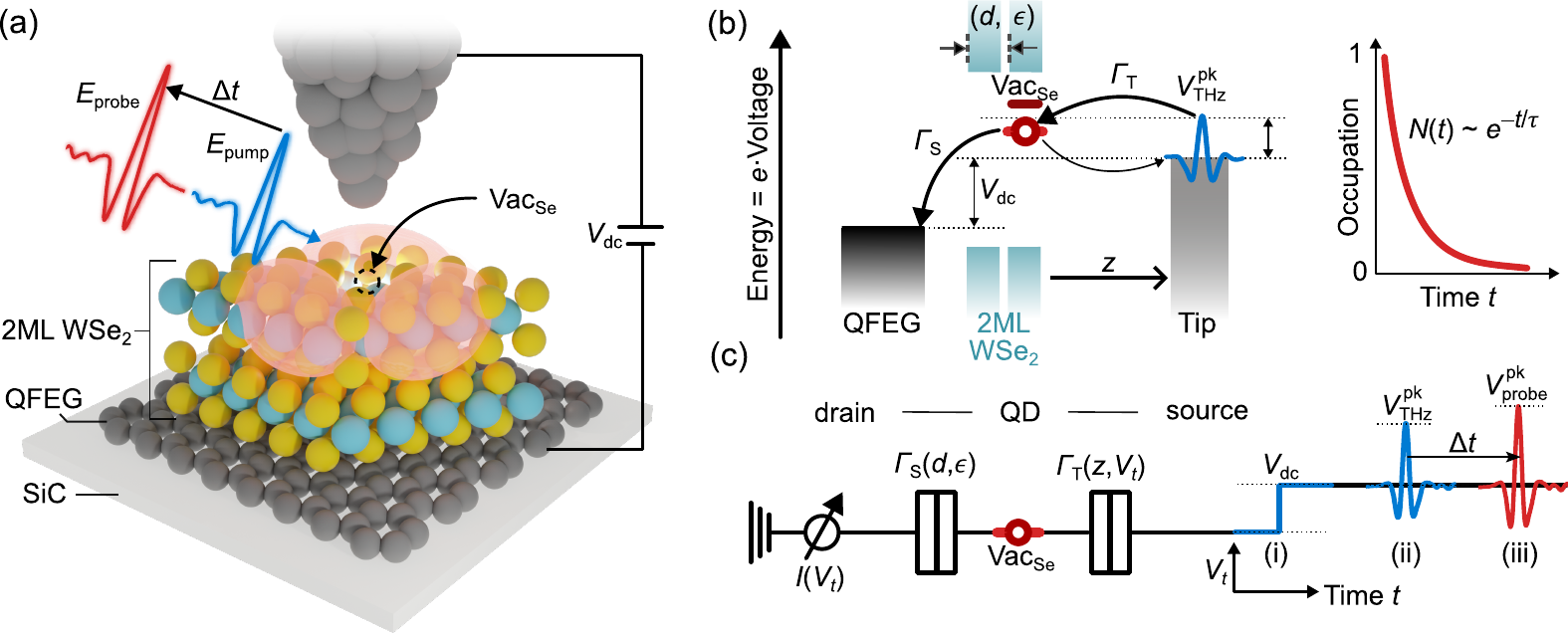}
\caption{\label{fig1}\textbf{A selenium vacancy (\vacse) in few-layer \wse{} as a single atom quantum dot.} (a)~Illustration of the experiment: \vacse{} in two monolayer (2\,ML) \wse{} on quasi free standing epitaxial graphene (QFEG) and conductive SiC studied with lightwave-driven STM. Free-space propagating THz pulses induce transient voltage bursts at the STM tip that enable electron tunneling into localized defect states of \vacse. (b)~Energy diagram illustrating electron tunneling from the tip to the \vacse{} in a double-barrier model. The electron transfer depends on the different coupling strengths $\Gamma$ which are based on energy (voltage) differences, the tip--sample spacing $z$ and the \wse{} layer thickness $d$ that determines the coupling of \vacse{} to the tip and the substrate, respectively. The charge occupation at \vacse{} exponentially relaxes to its equilibrium state. (c)~Schematic representation of sequential electron tunneling depicted as an equivalent circuit. The tip voltage bias is the sum of: (i)~constant \vdc, and (ii)~single THz voltage pulse with peak amplitude \vthzpk, or (iii)~pump and probe THz pulses with individual amplitude and time delay $\Delta t$.}
\end{figure}
%%%%%%% FIGURE 1 %%%%%%%%

The STM geometry enables the resonant exchange of electrons ($e^-$) in the single atom \vacse{} QD between tip (source electrode) and sample (drain) through sequential tunneling processes (Fig.~\ref{fig1}a,b). A finite tunneling current occurs if the $e^-$ exchange proceeds in series, i.e. from the source (tip) to the QD to the drain (substrate), or vice versa. Thereby, the coupling strength to the source and drain can be precisely controlled via the tip--sample distance and TMD layer thickness, respectively. 
In standard STM/STS, static lead potentials enable measuring energy selective, stochastic tunneling processes and average QD populations. Ultrafast pump-probe schemes in LW-STM allow us to access also the charge dynamics of the atomic QD. In the near-field of the tunneling junction, strong-field THz source pulses act as ultrafast bias transients that add to any static dc bias, as illustrated schematically in the equivalent circuit diagram in Fig.~\ref{fig1}c. 
We preserve state-selectivity of the tunneling process by tailoring the amplitude
and carrier-envelope phase of the THz transient to a single-cycle, unipolar pulse resonant with an electronic defect state~\cite{cockerTracking2016,mullerImaging2023,bobzienUltrafast2024}. \\

The key observable in LW-STM is the rectified charge \qlw{} transferred between tip and substrate per lightwave (THz) transient. Similar to the current in conventional STM, \qlw{} can be efficiently increased by reducing the tip--sample distance $z$.
In a QD geometry, a rectified charge requires two consecutive tunneling events from one lead to another via the discrete QD state. Here, two regimes can be distinguished: $Q_\text{LW}\leq 1e/$pulse and $Q_\text{LW}>1\,e/$pulse. In the latter regime, multiple $e^-$ tunnel per source pulse, which is only possible if the state that facilitates tunneling is short-lived, indicating strong QD--substrate coupling, $\Gamma_\text{S}$, where the charge-state lifetime ($\tau_0$) is much shorter than the source pulse duration ($\tau_\text{LW}$), $\tau_0 = \Gamma_\text{S}^{-1} \ll \tau_\text{LW}$. 
The coupling of the tip electrode can be tuned continuously from the weak-injection regime, ($\Gamma_\text{T}^{-1} \gg \tau_\text{LW}$) to the strong-injection regime. In van der Waals materials, and in particular for localized in-gap states, the weak coupling between layers and across 2D heterostructures results in long-lived states.
As a result, the regime of $Q_\text{LW}>1\,e/$pulse is not always accessible, because the large Coulomb energy associated with double charging of the QD inhibits multiple electron transfers within the short pulse duration. However, by using two time-delayed THz pulses it becomes possible to probe $e^-$--$e^-$ correlations, such as the Coulomb blockade in the time domain. \\

Previous studies could show that the average charge-state lifetime $\tau_0$ can be estimated via the maximum tunneling rate at saturated current through a localized state~\cite{kaiserChargestate2023,steurerLocal2014}. Specifically, when the tip--sample coupling $\Gamma_\text{T}$ greatly exceeds the substrate coupling $\Gamma_\text{S}$, the relationship $\Gamma_\text{tot}^{-1}= \Gamma_\text{S}^{-1} + \Gamma_\text{T}^{-1} \approx \tau_0$ holds, which can be achieved by approaching the tip. Additional \wse{} layers effectively decouple \vacse{} and substrate, allowing control over $\tau_0$ across several orders of magnitude. In a previous study, we reported average charge-state lifetimes of single \vacse{} in the top layer of 1-4\,ML \wse{}, ranging from 1\,ps to several ns~\cite{bobzienLayerDependent2024}. Owing to the spatial distribution of the defect orbital, top and bottom \vacse{} in the same layer exhibit similar coupling strength to the substrate and thus comparable charge-state lifetime~\cite{bobzienLayerDependent2024}. This study lays the basis for investigating ultrafast charge dynamics in the time domain using LW-STM.
As a starting point, we consider a neutral \vacse{} in 1\,ML and 2\,ML \wse{} characterized by the LUMO and LUMO+1 states (Fig.~\ref{fig2}a), residing in the unoccupied spectrum of the \wse{} band gap. STM topography and orbital images reveal the three-fold symmetric orbital shape of the defect state (Fig.~\ref{fig2}b)~\cite{schulerLargeSpinOrbitSplitting2019}. 

\section{Quenching of lightwave-driven orbital imaging}\label{sec3} \label{sec3}
%%%%%%% FIGURE 2 %%%%%%%%
\begin{figure}[t]
\centering
\includegraphics[width=0.7\textwidth]{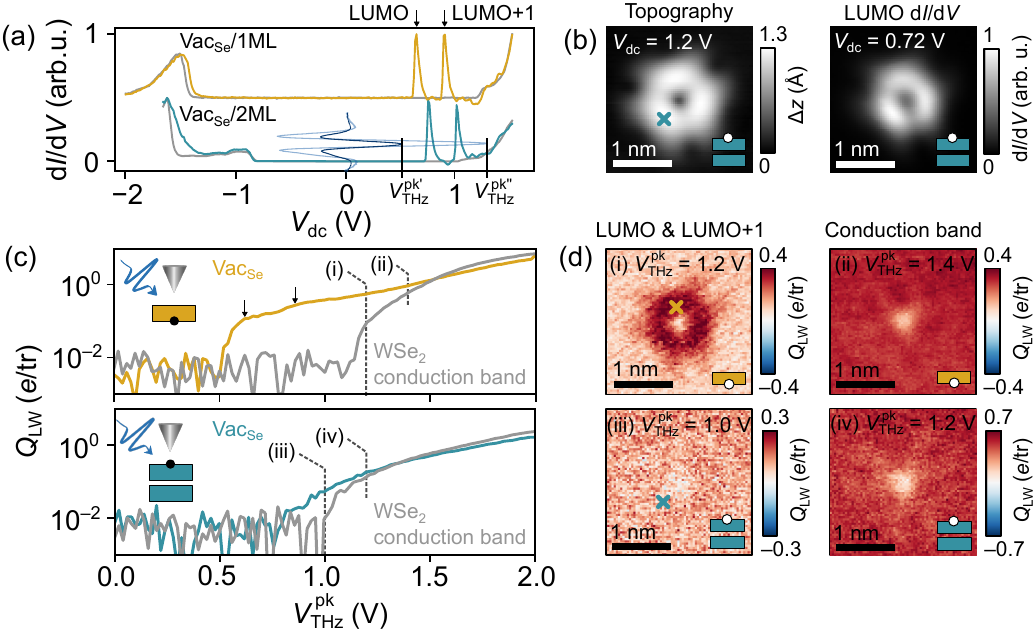}
\caption{\label{fig2}\textbf{\vacse{} defect states accessed via ultrafast LW-STS and LW-STM.} 
(a)~\didv{} spectra of \vacse{} in 1\,ML (yellow) and 2\,ML (turquoise) referenced to pristine \wse{} (grey). (b)~STM topography and constant-height \didv{} orbital mapping of a top \vacse{} in 2\,ML \wse. The height set point $z_0$ corresponds to \vdc$=1.2$\,V and $I_\text{dc}=100$\,pA (Methods). (c)~LW-STS of \vacse{} in 1\,ML (yellow) and 2\,ML (blue) each referenced to pristine \wse{} (grey). The tip--sample distance is $z_0-4$\,Å for 1\,ML and $z_0-3$\,Å for 2\,ML. (d)~LW-STM orbital images of \vacse{} in 1\,ML (upper panels) and 2\,ML (lower panels) at THz peak voltage indicated with (i)-(iv) in panel~(c). All lightwave measurements are performed at \vdc$=0$\,V. Integration time per data point is 2\,s for LW-STS and 20\,ms for LW-STM, corresponding to 20\,M and 0.2\,M THz pulses at 10\,MHz repetition rate.}
\end{figure}
%%%%%%% FIGURE 2 %%%%%%%%

To maximize the time resolution of THz source pulses and minimize dc background currents, the majority of LW-STM/STS experiments have so far been performed at or close to $V_\text{dc}=0$\,V. 
Fig.~\ref{fig2}c shows the rectified charge per THz waveform (transient) in units $e$/tr as a function of peak THz voltage, which in first order reproduces the $I(V)$ characteristic of the tunnel junction \cite{roelckeUltrafast2024}. The onset of the two rising edges in the $Q_\text{LW}(V^\text{pk}_\text{THz})$ curve (black arrows) align with the expected spectral locations of the lowest unoccupied molecular/defect orbital (LUMO) and LUMO+1. Additionally, the LW-STM map of the LUMO [Fig.~\ref{fig2}d (i)] closely resembles the dc STM measurement. However, the situation is strikingly different for LW-STS of \vacse{} on 2\,ML \wse{} (Fig.~\ref{fig2}c turquoise). In this case, we are unable to resolve the defect states through LW-STM [Fig.~\ref{fig2}d (iii) and (iv)], even though their STS spectrum matches closely the spectrum of \vacse{} on 1\,ML \wse{}. Instead of observing an enhancement in the LW-driven rectification at the defect position, which would typically arise from additional tunneling channels, we see a reduction of LW-driven tunneling at the defect center. This observation suggest that the \vacse{} on 2\,ML \wse{} enters a different lifetime regime, affecting the ultrafast tunneling process. The sub-nm point-like reduction of LW rectification to the conduction band in configurations (ii) and (iv) differs from the defect orbital and appears as a consequence of local band bending due to the charged \vacse{}~\cite{schulerHowSubstitutionalPoint2019,aghajanianResonant2020}.\\

We attribute the loss of in-gap defect state contrast in 2\,ML to an enhanced charge-state lifetime, related to the electronic decoupling of \vacse{} via the additional \wse{} layer \cite{bobzienLayerDependent2024}. The average charge-state lifetime of different \vacse{}/2\,ML obtained from dc current saturation measurements ranges from $\tau_0^\text{2\,ML} = 50$\,ps to 86\,ps (Extended Fig.~\ref{SI:fig:defect_characterization}), significantly longer than the sub-1\,ps source voltage duration. Consequently, multiple tunneling events per pulse are highly unlikely. Subsequently to the injection by the source pulse, the $e^-$ can relax through two possible pathways: (i) forward tunneling to the substrate, contributing to a measurable current, or (ii) backward tunneling to the tip.
Due to the small duty cycle in LW-STM, intermediate to strong charge injection from the tip is required to facilitate tunneling within the short duration of the  THz peak field. Hence, back tunneling is the more probable pathway if the charge-state lifetime exceeds a few picoseconds, quenching charge rectification.
As a result, a transiently charged \vacse{} in 2\,ML is more likely to relax by discharging via the STM tip (back tunneling) instead of the substrate (forward tunneling), resulting in zero net charge rectification. In contrast, the smaller lifetime of \vacse{} in 1\,ML ($\tau_0^\text{1\,ML} < 3$\,ps) enables notable charge rectification from tip to substrate during the transient THz pulse, which explains the orbital contrast observed in the LW-STM images for \vacse{} in 1\,ML \wse{}. For THz peak voltages reaching the conduction band, femtosecond carrier relaxation and strong delocalization of continuum states enable sub-cycle charge rectification during the THz gate irrespective of the layer thickness, yielding an unperturbed onset of conduction band states (grey curves in Fig.~\ref{fig2}c).\\

\section{Back tunneling suppression via Franck--Condon blockade}\label{sec4}

%%%%%%% FIGURE 3 %%%%%%%% 
\begin{figure}[t]
\centering
\includegraphics[width=1.0\textwidth]{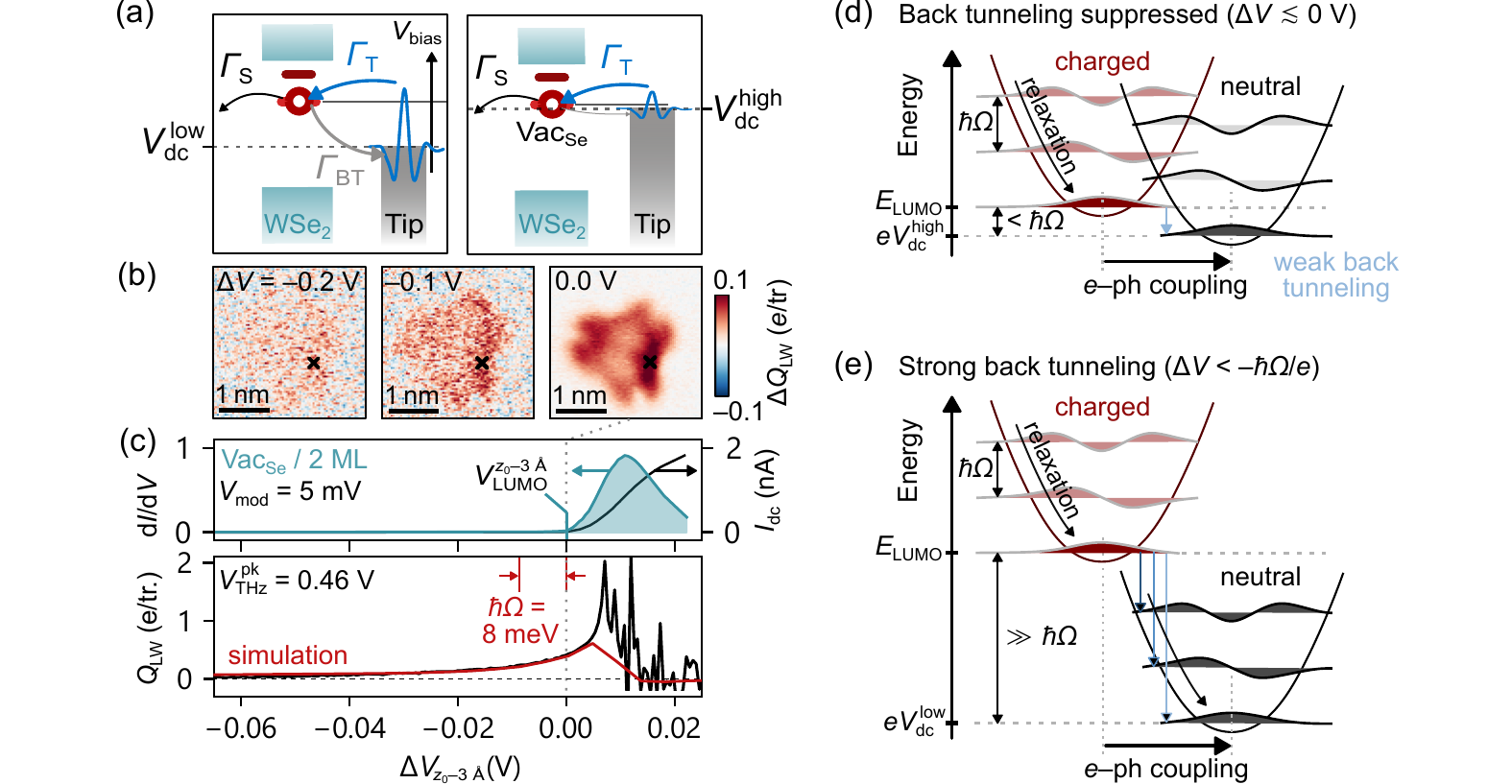}
\caption{\label{fig3} \textbf{Suppression of back tunneling via the Franck--Condon blockade.}
(a) Schematic illustration of reduced back tunneling rate at elevated \vdc. (b) LW-STM orbital images of \vacse{} showing a strong nonlinear increase of net rectified charge and orbital sensitivity as $\Delta V = V_\text{LUMO}-V_\text{dc}\rightarrow0$\,V. Measurement obtained at \vthzpk$=0.49$\,V, $V_\text{LUMO} = 0.7$\,V, and $z=z_0-3$\,\angstrom{}. (c) d$I$/d$V$ and $I(V)$ spectrum (upper panel) and rectified charge (lower panel) as a function of \vdc{} measured with \vthzpk$=0.32$\,V and $z=z_0-3$\,\angstrom{}. The simulation assumes a single vibrational mode $\hbar \Omega=8$\,m$e$V, a Huang-Rhys factor of 2.2, and experiment values for THz waveform, LDOS and $z$ (Methods). The data for panels (b) and (c) was obtained on Vac$_\text{Se}^{2.2}$ and Vac$_\text{Se}^{2.4}$ respectively (Extended Fig.~\ref{SI:fig:defect_characterization}). Black crosses in panel b indicate an equivalent position of the point spectrum. Integration time per data point is 100\,ms for LW-STS and 15\,ms for LW-STM, corresponding to 4.1\,M and 0.6\,M THz pulses at 41\,MHz repetition rate. (d) Transitions between vibronic states of the charged and neutral \vacse{} LUMO in the Franck--Condon picture at high \vdc{}($\Delta V \rightarrow 0$\,V) are limited to ground state transitions. (e) At reduced \vdc{} transitions to higher vibronic modes of the neutral state become available, enhancing the effect of back tunneling.}
\end{figure}
%%%%%%% FIGURE 3 %%%%%%%%

Next, we study the effect of the dc bias on THz charge rectification for \vacse{} in 2\,ML (Fig.~\ref{fig3}a).
To ensure comparability across different \vacse{}, we define a relative voltage offset from the LUMO $\Delta V_z = V_\text{dc} - V_\text{LUMO}(z)$ to account for local shifts of the relative LUMO position with tip height, lateral tip position, and sample heterogeneity (Methods). The first two factors arise from the well-known voltage drop across a double-barrier tunneling junction~\cite{nazinTunnelingrateselectron2005}.
In contrast to Fig.~\ref{fig2}d~(iii), where we set $V_\text{dc} = 0\,\text{V}$, we start to observe orbital contrast in LW-STM as \vdc{} nears the LUMO resonance, as seen in Fig.~\ref{fig3}b. This effect becomes more pronounced as $\Delta V_z$ approaches zero. We attribute the reestablished LW-STM contrast despite long-lived defect states to a suppression of back tunneling. This suppression can be explained by the so-called Franck--Condon blockade~\cite{kochTheory2006}, which is more significant than the (slowly-varying) bias-dependent change of the tunneling barrier.
When \vdc{} exceeds the onset of the \vacse{} resonance ($\Delta V_z > 0\,\text{V}$), significant dc currents occur, eventually leading to the saturation of $I_\text{dc}$ and causing charge fluctuations that substantially increase the noise in the THz current (Fig.~\ref{fig3}c).\\

In the Franck--Condon blockade regime, changes in the phonon population of a defect can introduce an asymmetry between forward and backward tunneling. In Figs.~\ref{fig3}d,e the principle of the Franck--Condon blockade is schematically illustrated. Within the Franck--Condon approximation, each charge state of the defect is represented by a harmonic potential with a single vibrational mode of characteristic energy $\hbar\Omega$. Due to fast vibrational cooling in 2D semiconductors~\cite{nieUltrafastCarrierThermalization2014} (curved black arrow), charge-state transitions primarily occur from the vibrational ground state (dark red), centered around the relaxed nuclei positions. The displacement of the potential energy surface between different charge states increases with the electron-phonon coupling strength. As a result, electron transfer (vertical blue arrows) is generally accompanied by vibrational excitation. However, when the dc bias is set close (within the order of the vibrational energy) to the LUMO orbital energy (Fig.~\ref{fig3}d), the overlap with energetically accessible low-lying vibrational states of the neutral defect is greatly reduced~\cite{leturcqFranck2009}. In contrast, when the dc bias is much lower (Fig.~\ref{fig3}e), all vibrationally excited states remain accessible. Consequently, back tunneling can be significantly suppressed by raising the electrochemical potential of the tip.\\

We model the transient charging and discharging of the atomic defect by a time-dependent Markov process, described by the so-called Master equation, as detailed in the Methods section. This model treats the defect by three electronic states (one ground state and two charged states, LUMO and LUMO+1), as well as a single vibrational mode ($\hbar \Omega\approx 8$\,meV) with multiple quanta, and a Huang-Rhys factor of $S \approx 2.2$ for a \vacse{} in 2\,ML \wse{}. These parameters are extracted from experimental STS spectra and the average charge-state lifetime $\tau_0$ from STM approach curves (Extended Fig.~\ref{SI:fig:simulations}a-b). The transition matrix is expressed as the sum of the transition rates due to sample, tip, and phonon relaxation. To calculate the rectified charge we model the actual waveform of the THz transient (Extended Fig.~\ref{SI:fig:simulations}c). The simulations, shown in Fig.~\ref{fig3}c (red), qualitatively and quantitatively reproduce the steep increase of the THz current as a function of dc bias within tens of meV of the defect resonance. Importantly, the strong suppression of back tunneling at elevated bias is attributed to a reduction in the Franck--Condon matrix elements. The Franck--Condon blockade acts as an effective backflow valve, favoring unidirectional charge transport when charge-state lifetimes exceed the pulse duration, which is crucial for achieving charge rectification in LW-STM.

\section{Optimal tip electrode distance between weak and strong coupling regime}\label{sec5}
%%%%%%% FIGURE 4 %%%%%%%%
\begin{figure}[t!]
\centering
\includegraphics[width=0.6\textwidth]{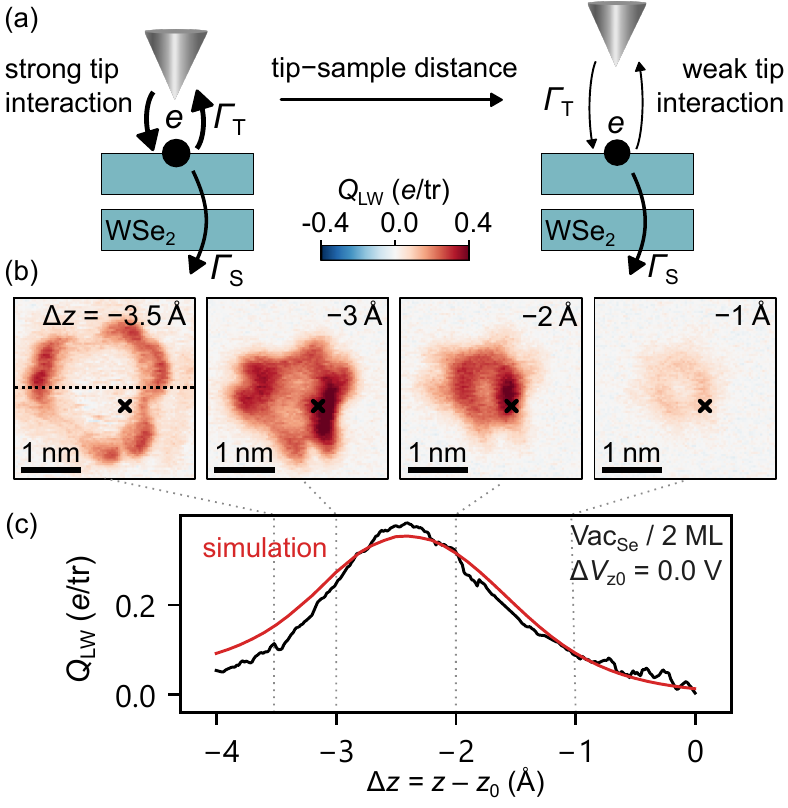}
\caption{\label{fig4} \textbf{Influence of strong and weak tip interaction on charge rectification.}
(a) Schematic illustration of the electronic coupling strength of \vacse{} to the STM tip at different tip height. (b) LW-STM orbital images with \vthzpk$=0.49$\,V and at \vdc$=0.7$\,V ($\Delta V_{z_0}=0.0$\,V). Images left to right at increasing tip--sample distance show a transition from the saturated ($\Delta z=-3.5$\,Å) to the unsaturated ($\Delta z=-1$\,Å) orbital shape. Reduced charge rectification at the center of \vacse{} at $\Delta z=-3.5$\,Å is related to strong back tunneling, i.e. efficient charge depletion by the tip. The dotted line marks the position for cross sections in Extended Fig.~\ref{SI:fig:orbital}b. (c) \qlw{} point spectrum (black curve) as a function of tip--sample distance measured on the orbital lobe of \vacse{}. The maximum a -2.5\,Å emphasizes the interplay of forward and backward tunneling, modeled by a simulation (red) that assumes a single vibrational mode with an energy of 8\,m$e$V and a Huang-Rhys factor of 2.2 (Methods). The data for panels b and d was obtained on Vac$_\text{Se}^{2.2}$ and Vac$_\text{Se}^{2.4}$ respectively (Extended Fig.~\ref{SI:fig:defect_characterization}).
Black crosses in panel b mark the equivalent position of the point spectrum in (c). Integration time per data point is 100\,ms for LW-STS and 15\,ms for LW-STM, corresponding to 4.1\,M and 0.6\,M THz pulses at 41\,MHz repetition rate.}
\end{figure}
%%%%%%% FIGURE 4 %%%%%%%%

The coupling of the atomic QD to the tip contact not only sensitively depends on the tip bias, but also on the tip--sample distance (Fig.~\ref{fig4}a). Fig.~\ref{fig4}b shows a series of LW-STM images illustrating the evolution of the orbital contrast at different tip heights at $V_\text{dc} = 0.7$\,V, corresponding to $\Delta V=0$ at $z_0$. The black curve in Fig.~\ref{fig4}c presents a $Q_\text{LW}$ approach curve obtained on a \vacse{} orbital lobe. 
Both, spatial (Fig.~\ref{fig4}b) and height (Fig.~\ref{fig4}c) dependence reveal an initial increase, followed by a decrease of the LW signal as the tip approaches the defect.
Initially, reducing the tip height compensates for the short duty cycle in LW-STM and enhances the signal. However, at very close distances, stronger back tunneling suppresses the rectified current, resulting in a maximum rectification near $z-z_0=-2.5$\,\AA. Two factors contribute to this effect: Due to the voltage drop across the \wse{} layers, a smaller tip height (or closer lateral tip position) shifts the LUMO resonance to higher voltages, thereby increasing $\Delta V$ and promoting back tunneling, as previously discussed. Additionally, the increasing tunneling rate $\Gamma_\text{T}(z)$ becomes much larger than the defect--graphene tunneling rate, $\Gamma_\text{S}$. This promotes back tunneling as the dominant discharging pathway, even for strong Franck--Condon blockades. 
Our simulations (Fig.~\ref{fig4}c, red) reproduce the shape and magnitude of the experimental $Q_\text{LW}(z)$ curve, accounting for both the $z$-dependent voltage drop and the increase in $\Gamma_\text{T}(z)$. While quenching of excited states via electron transfer to the tip was previously reported for photo-excited molecules~\cite{dolezalSingleMolecule2024,imai-imadaOrbitalresolved2022}, here we directly map this process with ultrafast time resolution.

\section{Time-domain detection of the ultrafast Coulomb blockade}\label{sec6}
%%%%%%% FIGURE 5 %%%%%%%%
\begin{figure}[t!]
\centering
\includegraphics[width=1.0\textwidth]{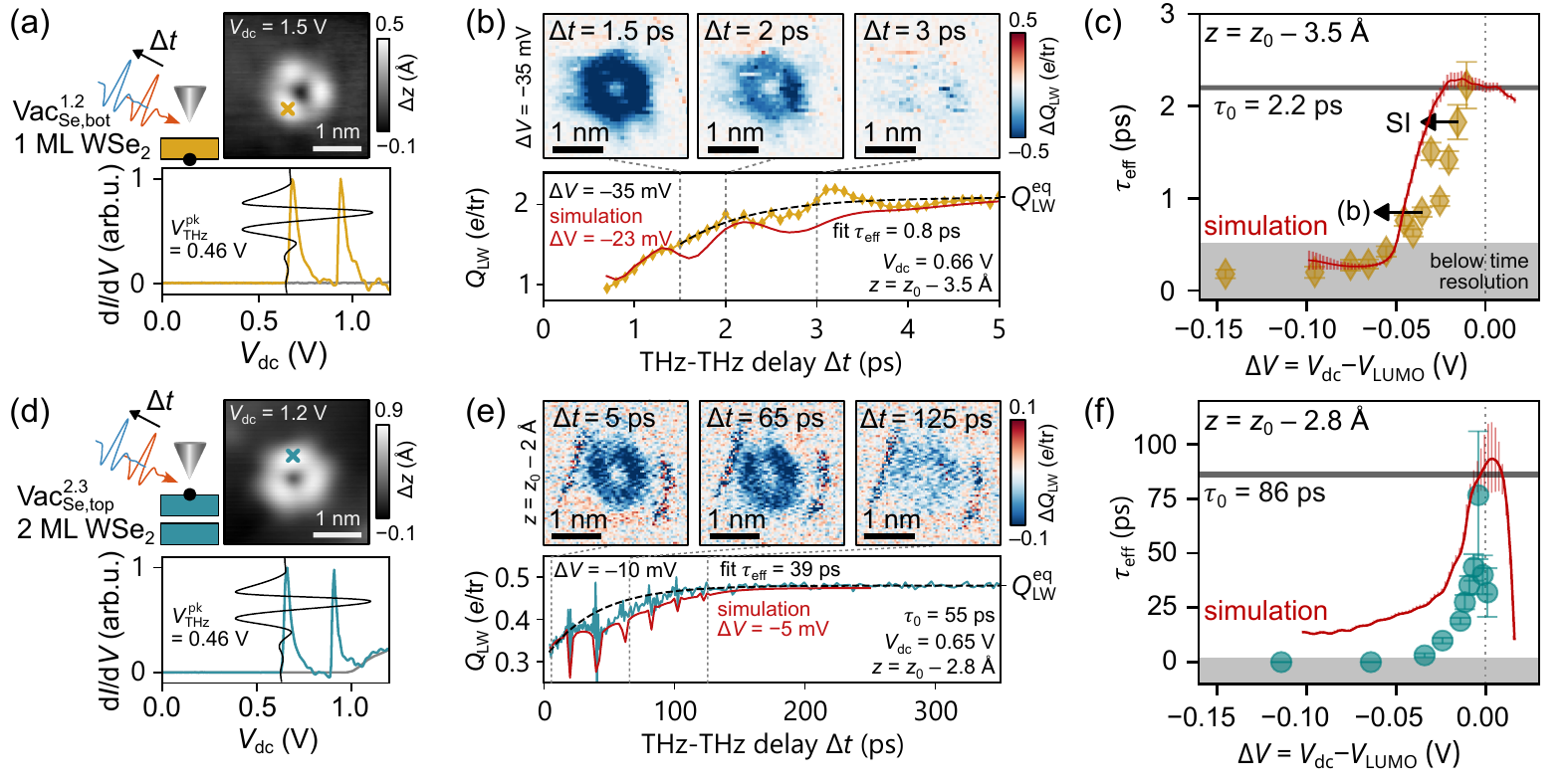}
\caption{\label{fig5} \textbf{Ultrafast Coulomb blockade detection using THz pump -- THz probe time-domain sampling.}
(a)~STM topography and STS spectrum of \vacse{} in 1\,ML \wse. The grey spectrum references pristine \wse. The illustration shows the concept of pump-probe measurements, where a THz pump pulse (red) charges the system while a time-delayed THz probe pulse (blue) measures the transient charge occupation. (b)~The rectified charge of the THz probe pulse as a function of THz-THz delay. The black line is an exponential fit to the relaxation of the charge state. Orbital images at selected delays, illustrating the change of rectified charge $\Delta Q_\text{LW}$ with respect to an equilibrium reference at 10\,ps. (c)~Dependence of the charge-state lifetime extracted via exponential fitting in panel b as a function of $\Delta V = V_\text{LUMO}-$\vdc. The horizontal line (grey) marks the average charge-state lifetime from dc measurements. Error bars correspond to the fit uncertainty and lifetime values below 0.5\,ps are shorter than the time resolution of the experiment. (d-f) Same analysis for \vacse{} in 2\,ML. Due to significant THz reflections, apparent as dips in $Q_\text{LW}(\Delta t)$, the extraction of a charge-state lifetime from the pump-probe measurement in panel (e) must be matched with a simulation as detailed in the Methods. The data was obtained on different \vacse{} characterized in Extended Fig.~\ref{SI:fig:defect_characterization}: Panels a-c at Vac$_\text{Se}^{1.2}$, panels d-e at Vac$_\text{Se}^{2.3}$, and panel f at Vac$_\text{Se}^{2.4}$. Integration time per data point is 100\,ms for LW-STS and 30\,ms (1\,ML) or 15\,ms (2\,ML) for LW-STM, corresponding to 4.1\,M and 1.2\,M or 0.6\,M THz pulses at 41\,MHz repetition rate.}
\end{figure}
%%%%%%% FIGURE 5 %%%%%%%%

Finding optimal parameters for $\Delta V$ and $z$ enables time-domain mapping of the transient Coulomb blockade close to its average charge-state lifetime. Here, we use a sequence of two time-delayed THz transients in combination with \vdc{}, such that \vdc+\vthzpk{} accesses the localized in-gap defect state, but remains below the onset of the conduction band (Fig.~\ref{fig5}a,d). In Fig.~\ref{fig5}b,e we show the transient Coulomb blockade at \vacse{} in 1\,ML and 2\,ML \wse{}.
In these measurements \qlw{}$(\Delta t)$ describes the rectified charge by the probe transient only (kHz amplitude modulation), while the pump transients interact at full repetition rate (41\,MHz). Near the overlap $\Delta t = 0$, the signal is dominated by interference between the THz waveforms, such that we restrict our analysis to time delays $\Delta t > 1.5$\,ps. For time delays much longer than the average charge-state lifetime $\tau_0$, the transient signal recovers to the equilibrium value $Q_\text{THz}^\text{eq}$, corresponding to unperturbed rectification by the probe transient. For orbital images acquired at different time delays, we plot the relative change of rectified charge $\Delta Q_\text{THz} = Q_\text{THz}-Q_\text{THz}^\text{eq}$, obtained by subtraction of a reference map at large delay ($\Delta t \gg \tau_0$). The differential images clearly reveal the orbital structure of \vacse{} with negative amplitude, marking the first movie of a Coulomb blockade at the atomic scale. \\

We model the lightwave-driven pump-probe (LW-PP) point spectrum by a single exponential fit (dashed black line) that allows to extract the relaxation dynamics of the \vacse{} quantum dot. This effective charge-state lifetime $\tau_\text{eff}$ combines the interplay of intrinsic charge relaxation to the substrate and ultrafast back tunneling to the tip. To disentangle these contributions, we record LW-PP spectra at increasing \vdc{}. We determine $\tau_\text{eff}$ via exponential fitting of the spectra at time delays between 1.5\,ps and 12\,ps, ensuring robust convergence of fit parameters. $\tau_\text{eff}$ strongly increases as $\Delta V \rightarrow 0$ and peaks near $\Delta V = 0$\,V, where it reaches the average charge-state lifetime $\tau_0$ (Fig.~\ref{fig5}c). At $\Delta V\sim0$\,V, back tunneling is strongly suppressed and the intrinsic relaxation channel to the substrate becomes the dominant mechanism for charge relaxation. Notably, the effective charge-state lifetime is rapidly quenched for negative $\Delta V$. This highlights the critical role of the Franck-Condon blockade for this experiment. At $\Delta V < -50$\,mV the time resolution and transient signal amplitude become insufficient to resolve the Coulomb blockade. Our master equation model enables simulations of charge rectification at various time delays using a model of our experimental waveform (Methods). Exponential fitting of simulated LW-PP spectra with similar boundary conditions as in the experiment clearly reveals the trend of increasing effective charge-state lifetime near $\Delta V=0$\,V (Fig.~\ref{fig5}c). \\

We performed the same series of measurements at a \vacse{} in 2\,ML \wse{} with an average charge-state lifetime $\tau_0 =86$\,ps.  
As the charge-state lifetime exceeds the time window of trailing reflections from windows in our beam path, which can amount up to $~30\%$ field amplitude (Extended Fig.~\ref{SI:fig:THz}), $Q_\text{LW}(\Delta t)$ deviates from a simple exponential decay, instead displaying dips indicative of interference with these reflections. Since $\Delta V$ values close to 0\,V are necessary to suppress back tunneling effectively, the impact of trailing reflections on the measurements can only be minimized, though not entirely eliminated. At elevated \vdc{} close to the onset of the defect state, the amplitude of THz pulse reflections is sufficiently high to interact with the \vacse{} states, causing additional excitations and experimental replica at various time delays.
Nevertheless, an ultrafast Coulomb blockade lasting over few-10\,ps is clearly revealed by the LW-PP spectrum and ultrafast orbital images in Fig.~\ref{fig5}e. For simple and robust extraction of an effective charge-state lifetime we determine the equilibrium signal at $\Delta t>250$\,ps and fit a single exponential model only to the unperturbed data from 1.5\,ps to 15\,ps. Simulated LW-PP spectra on basis of the Franck--Condon blockade and the full transient waveform with multiple reflections quantitatively model the experiment including the effect of THz pulse reflections (Methods, Extended Fig.~\ref{SI:fig:THz}). Limiting the fit interval to $\Delta t <15$\,ps causes large fit errors for $\tau_\text{eff}\gtrsim 20$\,ps, however, it ensures robustness of the fit in experiment and simulation. 
Again, the sharp decrease of $\tau_\text{eff}$ with more negative $\Delta V$ emphasizes the increasing relevance of back tunneling for \vacse{} in 2\,ML with extended $\tau_0$. 
As $\Delta V$ strongly depends on the lateral and vertical position of the tip, distinct $\Delta V$ regimes can coexist within a single LW-STM image. In the defect periphery, where $\Delta V > 0$, large dc currents permanently charge the defect, effectively quenching \qlw{}. At the circular transition region, where $\Delta V \approx 0$, charge fluctuations dominate, leading to significant noise in \qlw{} (Extended Fig.~\ref{SI:fig:orbital}c).\\

Previous studies of \vacse{} in \wse{} have investigated the coherent response of \vacse{} in 1\,ML \wse{} on Au(111) via LW-STM and LW-STS finding few-10\,meV dynamic shifts of the \vacse{} states upon transient charging \cite{roelckeUltrafast2024}. A coherent motion of the layer may be driven by inelastic excitations upon charging or discharching of \vacse{}. In contrast to fast charge relaxation on a metallic substrate, finite charge-state lifetimes and the Coulomb blockade in our samples impose a stochastically uncorrelated perturbation to any coherent signal since charge excitation and relaxation are not phase locked. At elevated \vdc{} the decoherence of lattice excitations is amplified by charge transfers from pulse replicas. We would expect an energy modulation of the defect state within a few 10\,mV driven by a coherent motion of the \wse{} layer as investigated by Roelcke \textit{et al.}~\cite{roelckeUltrafast2024} to also modulate the Franck--Condon blockade significantly. Our experiments and simulation indicate a coherent modulation (Fig.~\ref{fig5}b) that coincides with the THz pulse train measured also on pristine \wse{}. On basis of the present data and attributed to the incoherent nature of charge-state lifetime we cannot extract spectral evidence for a coherent motion of the \wse{} layer. Furthermore, the lack of a \wse{} moiré structure in our sample reduces mechanical strain and may facilitate fast delocalization of phononic exciations.

\section{Conclusions}
In conclusion, we demonstrate the direct observation of transient Coulomb blockade at an atomic quantum dot by lightwave-driven STM (LW-STM). Single selenium vacancies in monolayer and bilayer \wse{} on epitaxial graphene exhibit two distinct in-gap defect orbitals with charge-state lifetimes ranging from 2.2\,ps to 86\,ps due to the electronic decoupling. This platform provides an ideal testbed for studying ultrafast charge dynamics at the atomic scale. Our study examines the effect of varying coupling strength between the STM tip and the defect state, from weak to strong interaction. We identify back tunneling to the STM tip as a key challenge for LW-STM when accessing electronic states with charge-state lifetimes exceeding the pulse duration. 
However, quenching of the LW-STM signal can be mitigated by breaking the symmetry between charge transfer from and to the tip via the Franck--Condon blockade. The Franck--Condon blockade drastically reduces the charge-transfer rate from defect state to the tip by limiting the number of available vibrational transitions, acting as a charge backflow valve. By suppressing back tunneling, we achieve time-domain sampling of charge-state lifetimes close to their intrinsic values. A rate equation model accurately describes the time-dependent tunneling process across the different coupling regimes.
Using THz pump--THz probe time-domain sampling, we capture real-space snapshots of the transient Coulomb blockade. By fine tuning the transient THz waveform, dc bias offset, and tip height, we can deliberately populate specific defect orbitals and track the charge transfer into the substrate.
This work complements recent advances in observing defect resonances~\cite{jelicAtomicscale2024}, coherent lattice dynamics~\cite{roelckeUltrafast2024}, and charge density waves~\cite{shengTerahertz2024} with LW-STM, and establishes a pathway for probing ultrafast charge dynamics in low-dimensional materials. We envision that tunneling asymmetries induced by vibrational transitions, spin locking, or orbital angular momentum enable lightwave-driven charge control in complex materials at the frontier of ultrafast nanoscale electronics.

%%%%%%% METHODS %%%%%%%%

\section{Methods}

\subsection*{Sample}
Our \wse{} samples are grown via metal organic chemical vapor deposition on few-layer epitaxial graphene supported by a conductive silicon carbide (c-SiC) substrate. The plasma frequency of c-SiC is sufficient to effectively suppress THz reflections from the bottom surface. Intercalation of the graphene layers with hydrogen prior to \wse{} growth terminates the dangling silicon bonds, providing homogeneous electrostatic background. Further details of the sample growths procedure were previously reported along references \cite{bobzienLayerDependent2024, briggsInvitedRealizing2D2016}. Key advantages of epitaxially grown samples on quasi free-standing graphene (QFEG) are the electronic homogeneity (no moiré), a Fermi energy in the center of a large band gap, and weak hybridization with the substrate~\cite{bobzienLayerDependent2024}. The strong out-of-plane confinement facilitates electron-phonon coupling, well defined defect states, and high contrast of the defect states inside the band gap enabling LW-STM at nonzero \vdc. \\

High temperature annealing in UHV is required to clean samples after ambient exposure. We generate isolated \vacse{} in the top \wse{} layer via light sputtering with Ar ions approximately $2$\,s at $0.12$\,kV acceleration bias, $0.16$\,kV discharge voltage, and a sub-$\mu$A sputter-current~\cite{bobzienLayerDependent2024}. Multiple \vacse{} in 1-2\,ML \wse{} were investigated for this study, emphasizing the robustness and repeatability of the technique. Extended Fig.~\ref{SI:fig:defect_characterization} lists STM topography, d$I$/d$V$ spectra, and current saturation curves of relevant \vacse{}.

\subsection*{STM height set point and average charge-state lifetime}
For a reproducible definition of the tip height $z_0$, the STM feedback loop was opened on pristine \wse{} at a set point current of $100$\,pA with  applied bias voltage of $1.5$\,V (1\,ML) or $1.2$\,V (2\,ML), corresponding to an energy $200$\,meV above the conduction band onset. We estimate an absolute tip--sample distance at $z_0$ of $(7\pm2)$\,\angstrom. 
The average charge-state lifetime ($\tau_0$) of \vacse{} states scales with the decoupling layers of the sample and is determined via the current saturation in STM approach curves \cite{steurerLocal2014,kaiserChargestate2023,bobzienLayerDependent2024}. For simplicity we consider only the first unoccupied defect state (LUMO) of \vacse{} for the determination of $\tau_0$. Reduced localization due to a lower binding energy slightly reduces the average charge-state lifetime of the LUMO+1 defect state \cite{bobzienLayerDependent2024}. For different \vacse{}, $\tau_0$ slightly vary due to differences in their dielectric environment, however, the energy spectra and orbital shape remain comparable within few-10\,meV shifts (Extended Fig.~\ref{SI:fig:defect_characterization}).

\subsection*{Voltage drop and definition of $\Delta V$}
In our double barrier tunneling junction geometry, the effective voltage from the tip to the \vacse{} defect states depends on the \wse{} dielectric screening and the absolute height set point $z_0$. The relative voltage $\Delta V = V_\text{dc} - V_\text{LUMO}(z)$ is important to compare LW-STM data across different \vacse, where the onset of the LUMO, $V_\text{LUMO}(z)$, varies on the order of tens of m$e$V due to inhomogeneity of the sample and as a function of $z$. While for $\Delta V<0$ the dc current is negligible, $\Delta V>0$\,V generates a notable dc current and \vacse{} population in the experiment. In orbital imaging, we define $\Delta V$ at a single point on the orbital lobe of \vacse{}, avoiding complications of lateral dependence. 
We determine $V_\text{LUMO}(z)$ directly from a d$I$/d$V$ measurement at $z$ or via extrapolation of $V_\text{LUMO}(z_0)$ on the basis of measurements performed in Ref.~\citenum{bobzienLayerDependent2024}, Fig.~2. Owing to the high sensitivity towards thermal drifts on the sub-\angstrom{} scale, we estimate an accuracy $\delta_{\Delta V}\approx 10$\,mV.

\subsection*{LW-STM setup and near-field waveform detection}
Our LW-STM system is a commercial low-temperature STM from \textit{CreaTec Fischer \& Co. GmbH} with free-space optical access to the STM junction via large numerical aperture parabolic mirrors focused to the tip apex. THz pulses are generated via tilted pulse front optical rectification in lithium niobate pumped by a multi-MHz Yb:fiber laser with pump pulse energies of few µJ. The system achieves peak THz voltages at the STM tip up to 0.5\,V, 1\,V, and 2\,V at 41\,MHz, 20\,MHz, and 10\,MHz repetition rate respectively. We use frustrated internal reflection in a PTFE prism as well as a double or single mirror sequence to adapt the phase and polarity of single-cycle THz pulses at the STM tip \cite{allerbeckEfficient2023}. Pump and probe pulses are generated independently and recombined in collinear geometry via a c-cut sapphire window. Purging of the optics setup with dry air reduces ambient absorption before pulses are coupled into the vacuum chamber. Operating conditions of the scanning probe microscope are 5\,K base temperature and $10^{-10}$\,mbar pressure.\\

To discriminate lightwave-driven currents, we mechanically modulate the THz path at an intermediate focus using frequencies between 800\,Hz and \,1000\,Hz that are compatible with the $10^9$ transimpedance amplifier of the STM and show weak mechanical and electric noise. $I_\text{LW}$ is then extracted via lock-in demodulation and calibrated with respect to the absolute current. We convert current to rectified charge per transient ($e$/tr) on the basis of the laser repetition rate and the 50\,\% duty cycle of the modulation. In pump-probe measurements only the probe pulse is modulated, while pump pulses enter the junction at full repetition rate. We find that high dc currents prevent sensitive measurements of \qlw{} because of large noise floors.\\

The THz peak voltage is calibrated against \vdc{} using a recently developed approach based on the strongly nonlinear onset of the \wse{} conduction band \cite{bobzienUltrafast2024} measured in a clean pristine region of the sample (Extended Fig.~\ref{SI:fig:THz}a-c). On basis of this amplitude calibration we determine the relative time origin ($\Delta t = 0$) and temporal resolution of the experiment (380\,fs) by cross correlation of equal THz fields where only the sum field rectifies charges (Extended Fig.~\ref{SI:fig:THz}d). In a second step we measure the transient near-field waveform at the STM tip using THz cross correlation (THz-CC)~\cite{bobzienUltrafast2024,jelicAtomicscale2024}. Here, we set relative amplitudes such that a strong THz gate pulse rectifies electrons to the conduction band in a linear section of the \didv curve, while a time-delayed weak THz pulse modulates the THz peak voltage. This approach allows waveform characterization without requiring complex retrieval algorithms \cite{ammermanAlgorithm2022,bobzienUltrafast2024} (Extended Fig.~\ref{SI:fig:THz}e,f). Waveform sampling at extended delays reveals THz reflections at approximately 40, 60, 80, 100, and 120\,ps due to sapphire windows of the vacuum chamber, cryostat, and the beam combiner, which are unavoidable in our setup (Extended Fig.~\ref{SI:fig:THz}g). Thickness variation of sapphire windows and the beam combiner effectively reduces the amplitude of major THz reflection to below 30\,\% of the main pulse, causing reflections at 40 and 80\,ps to show distorted waveforms due to interference effects.

\subsection*{Theoretical Model}
The dynamics observed in the tunnel junction can be well represented by a time-dependent Markov process, which is described by the so called Master equation $\dot N = M \cdot N$, where $N$ is a vector containing the occupation probabilities of all considered states of the system and $M$ is the transition matrix.
Here, we considered three distinct electronic states: The electronic ground state at energy $E_0=0$, where the defect is charge neutral ($Q_0=0$), as well as two negatively charged states ($Q_{1,2}=-e$), with an additional electron in either the LUMO ($E_1$) or the LUMO+1 ($E_2$).
In addition, we model each electronic state with a single vibronic mode with 6 (14) vibrational excitation levels plus the vibrational ground state, yielding a total of 21 (45) different vibronic states for 2\,ML (1\,ML) \wse{}. The different number of required vibrational levels are due to different Huang-Rhys factors between 1\,ML and 2\,ML \wse{} (see below).\\

Transition between these states can occur due to inelastic charge transfer between defect and tip or sample, or via vibrational relaxation within one electronic state.
Accordingly, the transition matrix can be written as the sum of the transition rates due to sample, tip and phonon relaxation:
\begin{equation}
    M = M^\text{s} + M^\text{t} + M^\text{ph}
    \label{eq:matrix_components}
\end{equation}

The transitions related to a change of charge state are given by
\begin{align}
    M^\text{s}_{f\lambda^\prime i \lambda} &= T_{f\lambda^\prime i \lambda}(z_s, 0) \cdot V_{\lambda^\prime\lambda} \cdot \Upsilon_{fi} \\
    M^\text{t}_{f\lambda^\prime i \lambda} &= T_{f\lambda^\prime i \lambda}(z_t, V) \cdot V_{\lambda^\prime\lambda} \cdot \Upsilon_{fi}  \cdot \Lambda_{fi},
    \label{eq:TipSample_components}
\end{align}
with indices $i$ and $f$ representing the initial and final electronic state, $\lambda$ and $\lambda^\prime$ the initial and final vibrational state, $V_{\lambda^\prime\lambda}$ the Franck--Condon factors, and $T_{f\lambda^\prime i \lambda}$ the tunneling matrix element, respectively. $\Upsilon_{fi}$ captures the effect of various multiplicities of the involved states and $\Lambda_{fi}$ represents their relative coupling strength to the tip.\\

The tunneling matrix element $T_{f\lambda^\prime i \lambda}(z,V)$ describes the tunneling probability of an electron between tip or sample and the defect state, depending on bias voltag $V$ and distance z between defect and lead:
\begin{equation}
    T_{f\lambda^\prime i \lambda}(z,V) = 
        \begin{cases}
        \begin{aligned}
          & \int_{-\infty}^V \text{d}E\ g(E-\Delta E_{f\lambda^\prime i \lambda})\  e^{-2\kappa z} & \text{, if } Q_f - Q_i = -e\\
          & \int_{V}^{\infty} \text{d}E\  g(E+\Delta E_{f\lambda^\prime i \lambda})\  e^{-2\kappa z} & \text{, if } Q_f - Q_i = +e \\
          & 0\  & \text{, otherwise}
        \end{aligned}  
        \end{cases} 
    \label{eq:tunneling_matrix}
\end{equation}
with $\Delta E_{f\lambda^\prime i \lambda}$ being the energy difference between initial and final vibronic state and decay rate $\kappa=\sqrt{2m_e(\phi-E+eV/2)}/\hbar$.
The transition threshold is broadened by a Gaussian line shape $g(E)$ and normalized to integrate to the quantum of conductance $G_0=2e/\hbar$ for $z=0$.\\

For the Franck--Condon factors $V_{\lambda^\prime\lambda}$ we assume only a rigid shift in normal mode coordinates and negligible Duschinsky rotation~\cite{ZhebrakMethodForFranckCondon}:
\begin{equation}
% https://doi.org/10.1007/s10946-016-9552-1
    V_{\lambda^\prime \lambda} = \frac{\min(\lambda^\prime,\lambda)!}{\max(\lambda^\prime,\lambda)!}\ S^{|\lambda^\prime-\lambda|}\ e^{-S}\ \left [ L_{\min(\lambda^\prime,\lambda)}^{|\lambda^\prime-\lambda|}(S) \right]^2,
    \label{eq:franck_condon_elements}
\end{equation}
with Huang-Rhys factor $S$ and Laguerre polynomials $L_n^{\alpha}$.\\

The effect of the various multiplicities on the transition rates is captured by the factor $\Upsilon_{fi}$. Since we are only considering transitions for which either the final or initial state is a singlet, we can write $\Upsilon_{fi} = m_f$, with $m_f$ being the multiplicity of the final state.\\

The last term $\Lambda_{fi}$ for the tip-mediated transitions captures the relative coupling of LUMO and LUMO+1 to the tip, which can vary with lateral tip position. The additional factor $\Lambda_{fi}$ is normalized to 1 for the tip--LUMO coupling and has typical values between 0.5 and 1.0 for the tip--LUMO+1 coupling, chosen to match the experimental $\text{d}I/\text{d}V$ spectra.\\

For the charge-neutral phonon relaxations, the transition matrix is written as a decay to the vibrational ground state with lifetime $\tau_{ph}$:
\begin{equation}
    M^{ph}_{f\lambda^\prime i\lambda} = 
        \begin{cases}
        \begin{aligned}
          & 1/\tau_{ph} & \text{, if } f=i, \lambda^\prime=0 \text{ and } \lambda > 0 \\
          & 0  & \text{, otherwise.}
        \end{aligned}  
        \end{cases} 
    \label{eq:phonon_matrix}
\end{equation}
Lastly, the diagonal entries of the total transition matrix are set to $M_{ll} = -\sum_{k\neq l} M_{kl}$.\\

For static bias voltages, the master equation is solved for $\dot N = 0$ to obtain the bias and $z$ dependent equilibrium occupation $N_\text{eq}$. The dc current is then given by the net charge transfer between defect and sample $I_\text{dc} = \sum_k \left( W^s N\right)_k$, with $W^s_{f\lambda^\prime i \lambda} = M^s_{f\lambda^\prime i \lambda} \cdot (Q_f - Q_i)$.
In order to evaluate the evolution of the system under a transient bias voltage, the master equation is solved in discrete time steps $N_{i+1} - N_i = M\left(V(t_i)\right) \Delta t \cdot N_i$, starting from the equilibrium occupation at $V_\text{dc}$.

\subsection*{Implementation of simulation}
The simulations based on the master equation require experimental parameters that can be obtained from static measurements of the \vacse{} (Extended Fig.~\ref{SI:fig:defect_characterization}). We extracted the energies of localized defect states $E_1$ and $E_2$ via the zero-phonon lines and the tip coupling factor $L_{f,i}$ from the experimental \didv{} spectrum at the set point $z_0$. The average charge-state lifetime $\tau_0$, and the tunneling barrier $\kappa$ were obtained via a rate equation fit to the $I(z)$ approach curve~\cite{bobzienLayerDependent2024}. \\

The vibronic broadening of the defect state resonances was approximated by a single vibrational mode with $\hbar\Omega = 8$\,m$e$V and a Huang-Rhys factor of $S=2.2$ (2\,ML) for both charge states. 
We find that a Gaussian lineshape with 3\,meV bandwidth provides a good match with the experimental \didv{} (Extended Fig.~\ref{SI:fig:simulations}b), while being consistent with literature on Vac$_\text{S}$ in WS$_2$, observing a dominant phonon mode of $12\,$m$e$V and Huang-Rhys factor of 1.2~\cite{schulerLargeSpinOrbitSplitting2019}. Considering an average 40\% softening of the phonon modes in WSe$_2$ as compared to WS$_2$~\cite{zhaoLatticedynamicsmono2013}, we choose a phonon mode of $\hbar\Omega = 8$\,m$e$V, and fit the Huang-Rhys factor to the STS spectrum. For the simulations on 1\,ML \wse{} we cannot directly fit the vibronic satellite peaks of the LUMO resonance. Instead, we rescale the Huang-Rhys factor from 2\,ML \wse{} by a factor of about two, as previously found for C$_\text{S}$ in 1\,ML and 2\,ML WS$_2$~\cite{cochraneSpindependent2021}. Therefore we choose $S=5$ for \vacse{} in 1\,ML \wse{}, while we assume the phonon mode energy to remain constant, $\hbar\Omega = 8\,$m$e$V~\cite{cochraneSpindependent2021}.\\

Although lifetime broadening of a state with tens of picoseconds of lifetime would lead to a Lorentzian lineshape with few tens of $\mu$eV broadening, we justify this approximation by the presence of additional low-energy vibrational modes with high Huang-Rhys factor ($S \gg 1$), e.g. well known interlayer phonons for multi-layer TMDs~\cite{jeong_Coherent_2016}. These low-energy phonons impose an additional Franck--Condon blockade that explains difference between simulation and experiment at bias voltages overlapping with the LUMO resonance ($\Delta V >0$). In particular the breakdown of simulations at $\Delta V\approx 10$\,mV in Fig.~\ref{fig3}c emphasizes the limitation of the 1-phonon mode treatment. A thorough 2-mode treatment would require significant more computational resources without major benefits to our model.\\

The effective strength of the Franck--Condon blockade also depends on the phonon lifetime $\tau_{ph}$. We distinguish three regimes in the evolution of charge-state transitions between tip and defect: (I)~A voltage gate for charging, (II)~vibronic relaxation, and (III)~charge relaxation, sketched in Extended Fig.~\ref{SI:fig:FC}. The THz peak voltage \vthzpk{} of few-100\,mV enables transitions from the neutral ground state to high vibronic modes of the charged state. Vibronic relaxation in a 2D semiconductor with strong electron-phonon coupling  typically occurs faster than the THz gate timescale. Additionally, due to local resolution of LW-STM, combined with the fast lateral diffusion of phonon modes, the atomic-scale quantum dot remains unaffected by the lifetime of acoustic phonons. As a result, we expect relaxation of the charged \vacse{} to the vibrational ground state within few-100\,fs, as seen for thermalization and cooling of photoexcited charge carriers~\cite{nieUltrafastCarrierThermalization2014}. We conservatively assume $\tau_{ph}=1$\,ps for the simulations. After the source pulse and at small $\Delta V \lesssim 0$, charge-state transitions from low vibrational modes of the charged state have only few vibrational modes in the neutral state available, creating an asymmetry between forward and backward tunneling that reduces the impact of back tunneling. In addition, our simulation considers the LUMO and LUMO+1 multiplicity as a quartet (J = 3/2) and a sextet (J = 5/2) state, respectively, further enhancing this asymmetry (Fig.~S5 of Ref.~\citenum{schulerLargeSpinOrbitSplitting2019}).\\

The voltage drop in 2\,ML \wse{} is approximated by a plate capacitor model, $V_\text{eff} = V_\text{dc} \frac{z}{z+d/\epsilon_r}$, with a gap composed of vacuum and \wse{} using a dielectric constant $\epsilon_r=6.5$ and $d=6.5$\,\angstrom{} for \wse{} layer spacing~\cite{lin_Atomically_2014} (Extended Fig.~\ref{SI:fig:simulations}e). Due the complex tip shape and iso-potential surfaces that are not parallel to the substrate plane, the voltage drop not only depends on the tip height $z$ but also on the lateral position of the tip.
An accurate description of the effective bias potential can be obtained by integrating the wavefunction with the spatial dependence of the electric potential within the junction~\cite{kraneMapping2019}. In most simulations shown in this work, this effect is not crucial to reproduce the experimental data. However, for the $Q_\text{LW}(z)$ measurements this simple plate capacitor model is not sufficient to model the experiment quantitatively, see Extented Fig.~\ref{SI:fig:simulations}g (red line). Because the Franck--Condon blockade depends strongly on the effective applied \vdc{}, it is very sensitive to changes in the voltage drop and therefore needs a more sophisticated model for the voltage drop. For this reason, we introduced a slightly modified junction model for the $Q_\text{LW}(z)$ measurements. The tip position for tunneling is assumed to be at height $z_0$, whereas the effective plane of the plate capacitor is at larger distance from the sample $z_\text{VD} > z_0$ (Extended Fig.~\ref{SI:fig:simulations}e-g). For the 1\,ML \wse{} the simple plate capacitor model is used with an $d/\epsilon_r = 0.1$\,\AA.\\

To match the simulations of the \vacse{} in 1\,ML \wse{} in Fig.~\ref{fig5} quantitatively, we assume a 0.4\,\AA{} uncertainty on the experimental $z_\text{0}$ that may arise from small lateral variations in the tip position between the $I(z)$ spectroscopy and THz pump-probe measurements.\\

The simulation includes a model of the extended transient waveform to reproduce the experimental time-domain signal. Since pump and probe are indistinguishable in the experiment and equally interact with the \vacse{} state, the Coulomb blockade causes a transient signal that is symmetric at $\Delta t = 0$\,ps. This significantly impacts measurements at $\Delta t>15$\,ps, where transient signals from main pulse and reflections overlap causing a non-trivial curvature. Thus, we restrict the data interval for exponential fitting between 1.5\,ps to 15\,ps. As another consequence of reflections, extended coherent signals are quenched due to destructive interference from multiple excitations. 

%%%%%%% METHODS %%%%%%%%

\section*{Acknowledgements}
We acknowledge fruitful discussions with Oliver Gröning and generous support by Roman Fasel.
This research was funded by the European Research Council (ERC) under the European Union's Horizon 2020 research and innovation program (Grant agreement No. 948243). SEA and NK appreciate financial support from the Werner Siemens Foundation (CarboQuant). Funding for DCF and JAR is through the National Science Foundation EEC-2113864 and ECCS-2202280. For the purpose of Open Access, the author has applied a CC BY public copyright license to any Author Accepted Manuscript version arising from this submission.

\subsection{Data availability}
The data supporting this study is available upon reasonable request. 

\subsection{Code availability} 
Code is available from the corresponding author upon reasonable request.

\subsection{Author contribution}
$\dag$ These authors contribute equally to this work.\\
JA, LB, and BS conceived the experiment. JA, LB, and BS designed and implemented the setup for LW-STM. JA and LB performed the measurements. JA, LB, and NK analyzed the data. NK and SEA performed the modelling. DCF and JR grew the samples. JA, LB, NK, and BS wrote the manuscript with feedback from all authors.

\clearpage

%%==================================%%
%% EXTENDED FIGURES %%
%%==================================%%
\renewcommand{\thefigure}{S\arabic{figure}}
\setcounter{figure}{0}  

%adjust linespacing
\renewcommand{\baselinestretch}{1.2} 
\section{Extended Figures}

\begin{figure}[H]
\centering
\includegraphics[width=1.0\textwidth]{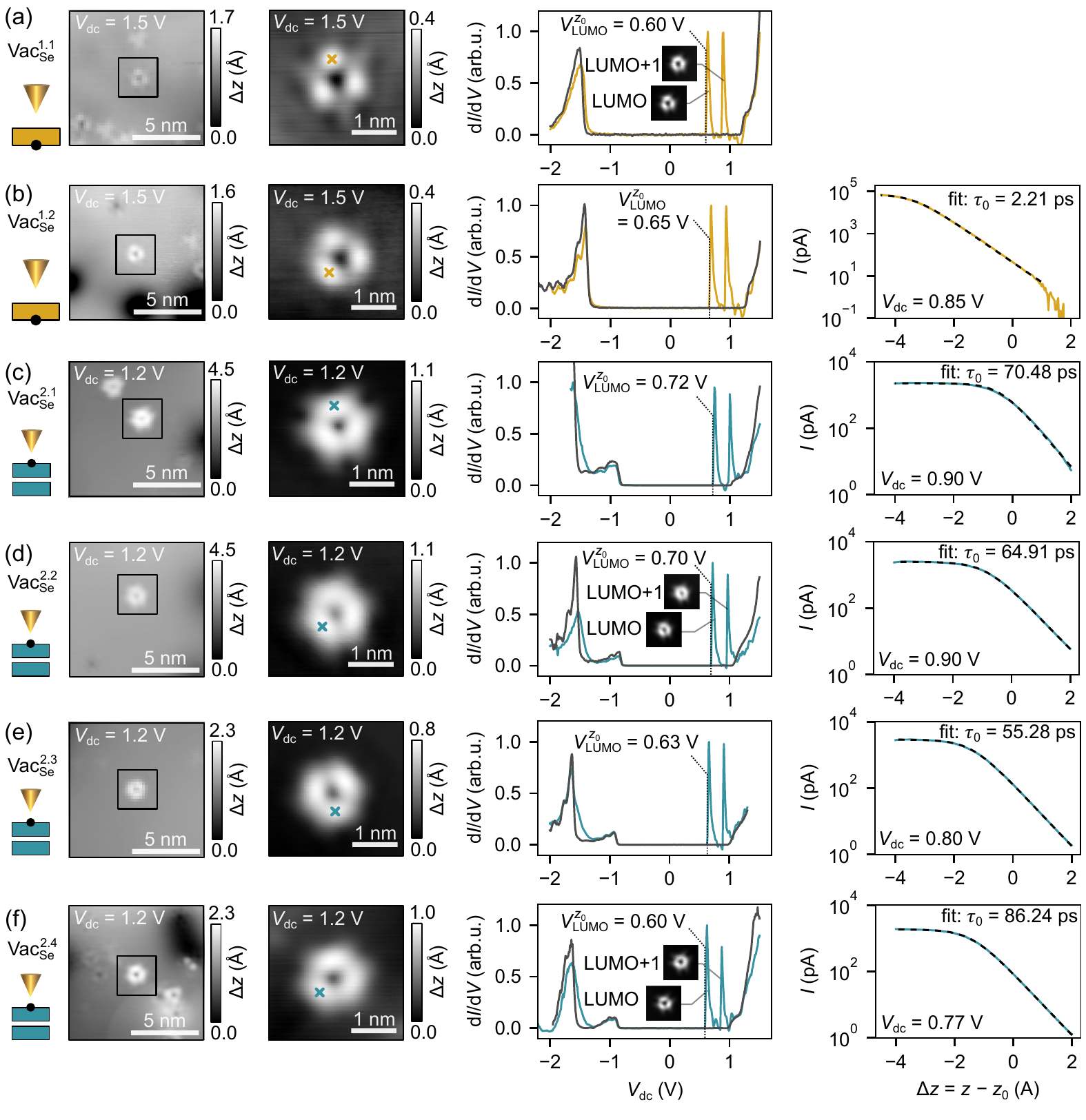}
\caption{\label{SI:fig:defect_characterization}\textbf{STM characterization of different Vac\textsubscript{Se} with unique identifiers in the name superscript.} Every row shows an overview and closeup STM topography, STS spectra of \vacse{} and pristine \wse, and STM approach curves. (a,b)~Isolated Vac\textsubscript{Se} in 1\,ML \wse{} in the bottom sulfur layer, identified by STM topography. Top and bottom vacancies of the same \wse{} layer are comparable regarding electronic structure and charge-state lifetime~\cite{bobzienLayerDependent2024}.
(c-f)~Various top \vacse in 2\,ML \wse. The average charge-state lifetime $\tau_0$ at the LUMO resonance is estimated via the saturation of \idc$(z)$ at \vdc above the LUMO and below LUMO+1. $\tau_0$ varies between vacancies varies due to the electrostatic background of the sample. Insets in the \didv spectra in rows~(a), (d) and (f) show constant-height \didv orbital maps of the LUMO and LUMO+1 resonances respectively. The bias modulation amplitude for all \didv measurements is $V_\text{mod}=10$\,mV.
}
\end{figure}

\begin{figure}[H]
\centering
\includegraphics[width=1.0\textwidth]{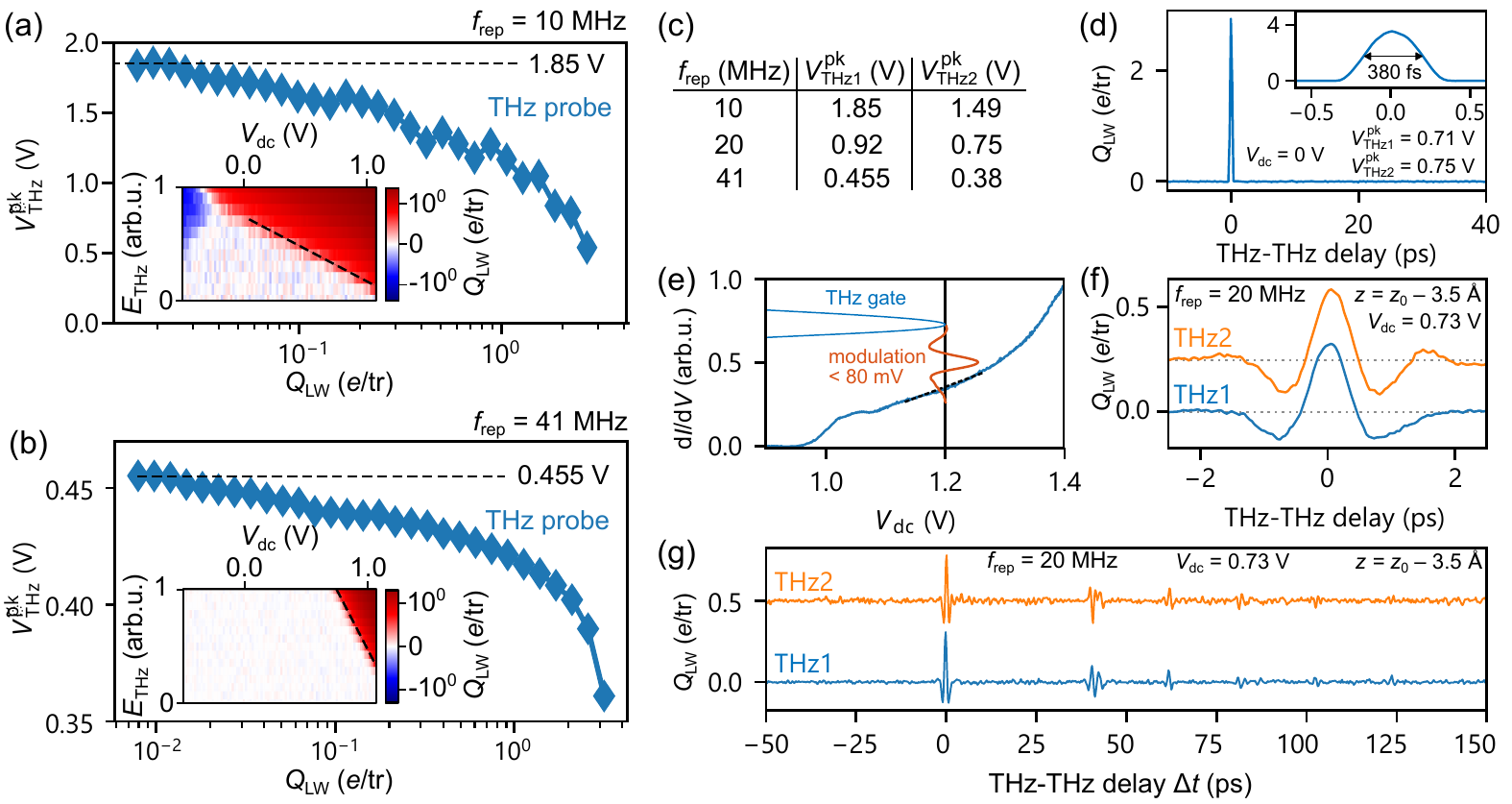}
\caption{\label{SI:fig:THz}\textbf{THz waveform and amplitude calibration on pristine 2\,ML \wse.} (a,b)~Amplitude calibration of the positive THz peak voltage of our unipolar THz pulses at the STM tip in tunneling contact with pristine \wse{} at 10 and 41\,MHz repetition rate. The insets shows the relevant section of THz field amplitude versus dc bias obtained in the experiment and used for calibration as discussed in reference~\cite{bobzienLayerDependent2024}. (c)~Table of THz peak values obtained for both THz arms at variable repetition rates. (d)~Rectified charge by THz pulse correlation using THz peak voltages that only in sum reach to the conduction band. The inset indicates the overlap region with a FWHM temporal width of 380\,fs. (e)~Illustration of THz-CC waveform sampling \cite{bobzienLayerDependent2024,jelicAtomicscale2024} performed in a linear region of the \didv curve. (f)~Transient voltage waveform of both THz arms, and (g)~extended delay time showing multiple reflections from sapphire windows of the vacuum chamber~\cite{allerbeckEfficient2023}. THz-CC measurements were performed at 20\,MHz and \vdc$=0.73$\,V, with a peak field of the THz gate $V_\text{THz,gate}^\text{pk} = 0.47$\,V and THz modulation $V_\text{THz,mod}^\text{pk}< 0.07$\,V    
}
\end{figure}

\begin{figure}[H]
\centering
\includegraphics[width=1.0\textwidth]{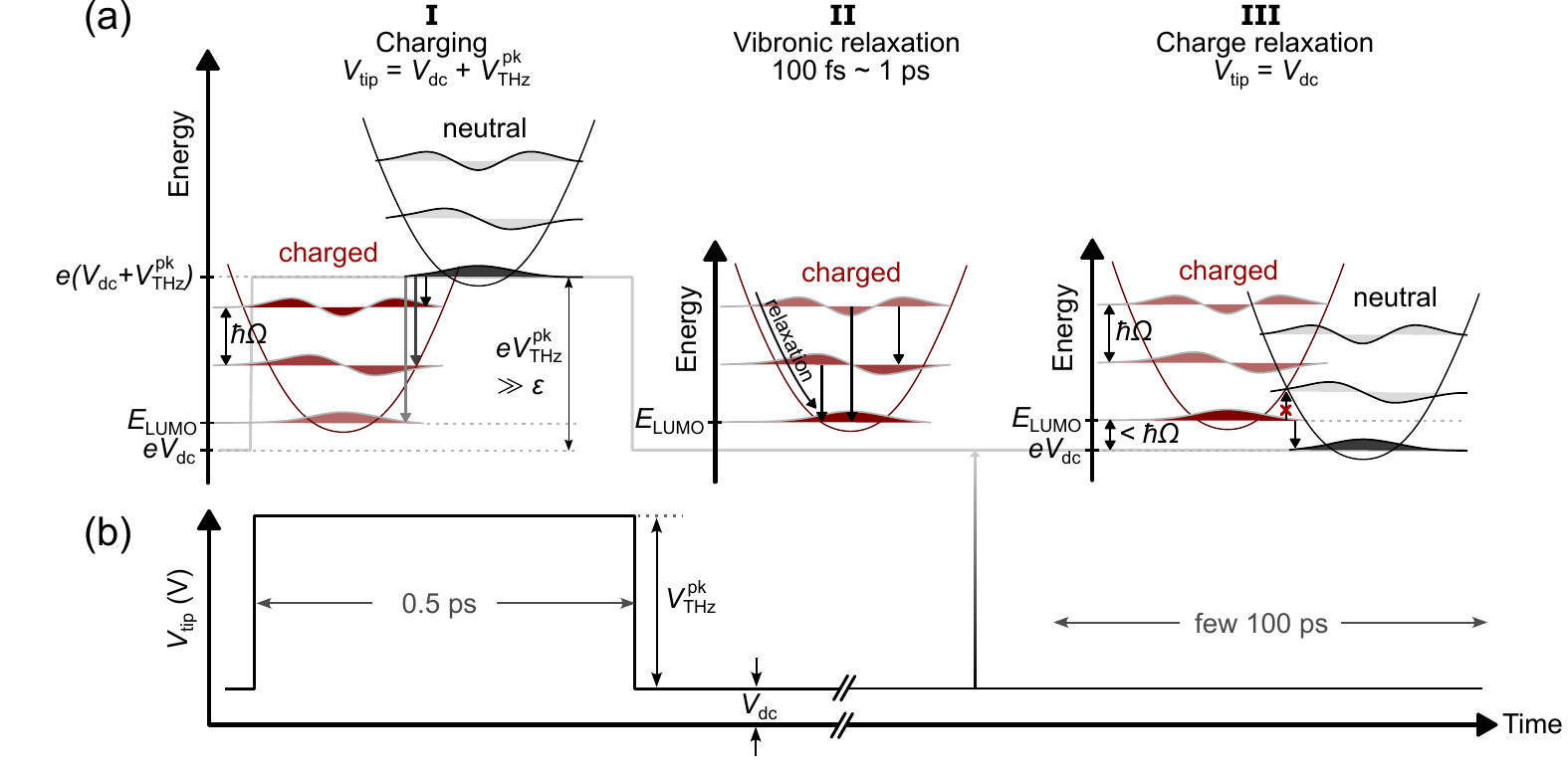}
\caption{\label{SI:fig:FC}\textbf{Suppression of back tunneling via the Franck--Condon blockade.} (a)~Illustration of charging, vibronic relaxation and back tunneling in a time-dependent Franck--Condon picture with three time steps. The neutral and charged states feature multiple quanta (levels) of a vibronic mode. (I)~During the voltage gate with \vthzpk$\gg\hbar\Omega$ all vibronic levels are energetically accessible, however, the wavefunction overlap may be enhanced for transitions to higher vibronic levels of the charged state. (II)~We estimate fast vibronic relaxation on few-100\,femtosecond to 1\,ps timescales, hence during or shortly after the THz voltage gate. (III) With high \vdc, back tunneling to higher vibronic levels of the neutral state is suppressed, compare to Fig.\,\ref{fig3}. (b)~Schematic time trace of a single rectangular voltage gate with 0.5\,ps duration that adds linearly to a constant \vdc. } 
\end{figure}

\begin{figure}[H]
\centering
\includegraphics[width=1.0\textwidth]{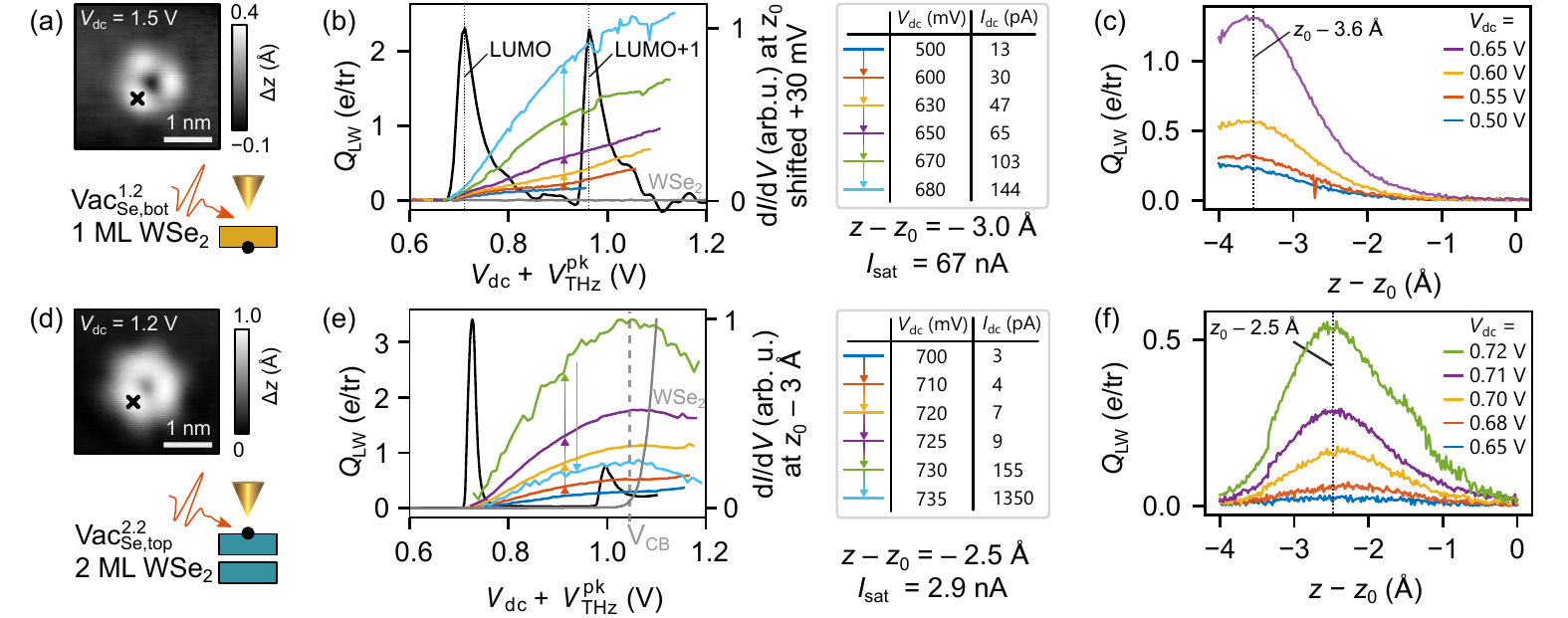}
\caption{\label{SI:fig:backtunneling}\textbf{THz amplitude spectroscopy and THz approach curves at \vacse{} in 1 and 2\,ML \wse.} (a)~STM topography of a bottom \vacse{} in 1\,ML \wse. The black cross marks the point spectra position. (b)~THz amplitude spectra showing rectified charge by a single transient as a function of total peak voltage at the tip (sum of \vdc{} and \vthzpk) for different \vdc{} taken at $z=z_0-3$\,\angstrom. A constant height \didv spectrum (black curve) at $z=z_0$ is overlayed for reference and shifted by +30\,mV to compensate for the reduced tip--vacancy bias at reduced tip height. Kinks in the \qlw{}(\vthzpk) curves at 0.97\,V indicate the LUMO+1 resonance. (c)~LW-STM approach curves at various \vdc. At high \vdc{} a characteristic maximum at -3.6\,\angstrom{} appears. (d-f)~Same series of measurements for \vacse{} in 2\,ML \wse. Notably, the rectified charge first increases with \vdc{} but collapses at high \vdc{} (light blue curve) due to Coulomb blocking by dc electron tunneling when $I_\textrm{dc}$ is close to saturation $I_\text{sat}$. The local maximum of \qlw{} in panels (c) and (f) results from the interplay of charging and back tunneling which has a strong spatial distribution discussed in Fig.~\ref{SI:fig:orbital}(b). The data was obtained on Vac$_\text{Se}^{1.2}$ [panels (a)-(c)] and Vac$_\text{Se}^{2.2}$ [panels (d)-(f)] respectively (Extended Fig.~\ref{SI:fig:defect_characterization}). Integration time per data point is 400\,ms for LW-STS and 100\,ms (1\,ML) or 200\,ms (2\,ML) for $z$-spectroscopy, corresponding to 16\,M and 4.1\,M or 8.2\,M THz pulses at 41\,MHz repetition rate.}
\end{figure}

\begin{figure}[H]
\centering
\includegraphics[width=1.0\textwidth]{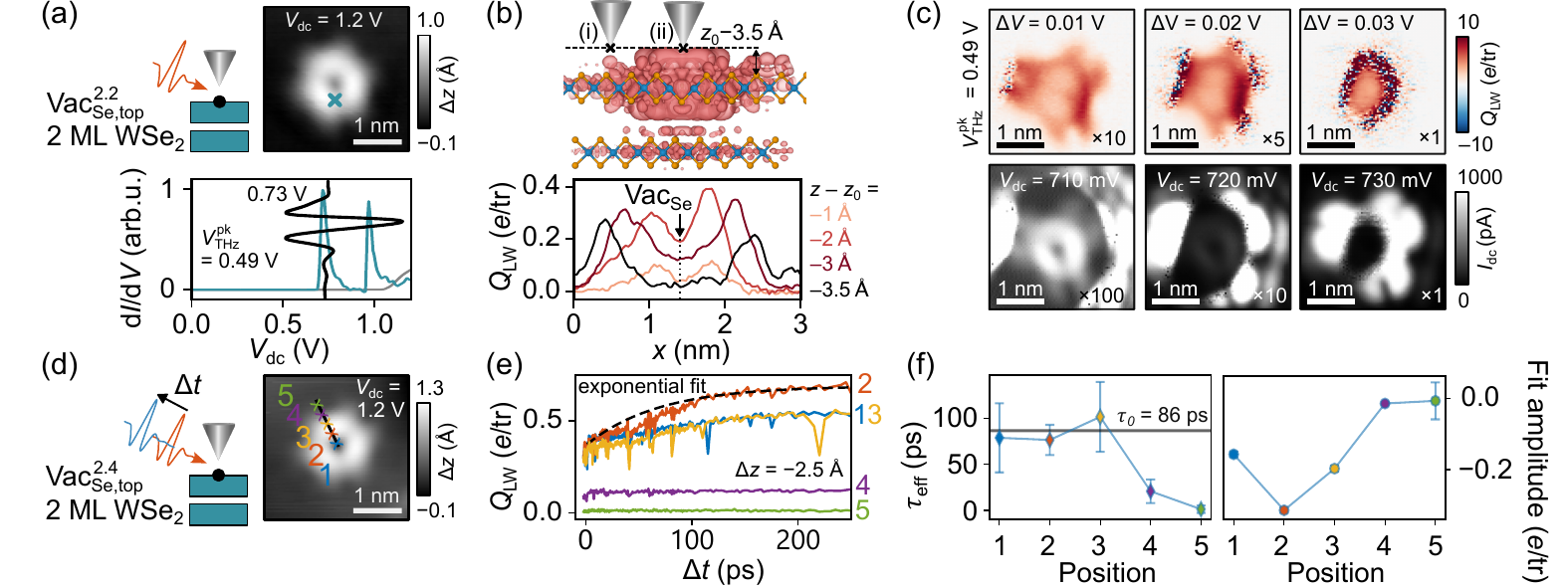}
\caption{\label{SI:fig:orbital}\textbf{Spatial dependence of the Coulomb blockade.} (a)~STM topography and \didv spectrum of \vacse{} in 2\,ML (same as in Fig.~\ref{fig5}d). (b)~Isosurface of the \vacse$_\text{top}$ orbital in 2\,ML \wse. The dashed line indicates the horizontal position of the STM tip in typical measurements with reduced tip--sample distance. Line profiles are extracted from Fig.~\ref{fig4}b of \vacse{} in 2\,ML \wse{} measured with LW-STM at \vdc$=0.7$\,V and different tip--sample distance. At reduced distance, back tunneling quenches the charge-state lifetime and hence also the rectified charge at the central part of \vacse{} leading to the ring-shaped orbital seen in Fig.~\ref{fig4}. (c)~LW-STM imaging (upper panels) and constant-height current maps at $\Delta V > 0$\,V and $z = z_0 - 3$\,\angstrom{} complementing the series in Fig.~\ref{fig3}b. At high \vdc, dc current maps reveal the saturated \vacse{} orbital in the periphery related to the voltage drop as a function of lateral distance. Fluctuations, seen as an increased noise in \qlw{} maps occur at the onset of \idc and are likely related to telegraph noise in the Franck--Condon blockade regime~\cite{kochTheory2006}. (d)~STM topography of a different \vacse{} in 2\,ML \wse. (e)~LW pump-probe measurements at different positions along the \vacse{} orbital indicated by numbers in panel (d). THz peak voltages are $0.18$\,V and $0.21$\,V for pump and probe pulses respectively. The height set point is $z = z_0 - 2.5$\,\angstrom{} and \vdc$=0.66$\,V. (f)~Charge-state lifetime and amplitude obtained from exponential fits to the data in~(e) with identical boundary conditions as used in Fig.~\ref{fig5}f. The data was obtained on Vac$_\text{Se}^{2.2}$ [panels (a)-(c)] and Vac$_\text{Se}^{2.4}$ [panels (d)-(f)], as characterized in Extended Fig.~\ref{SI:fig:defect_characterization}.}
\end{figure}

\begin{figure}[H]
\centering
\includegraphics[width=1.0\textwidth]{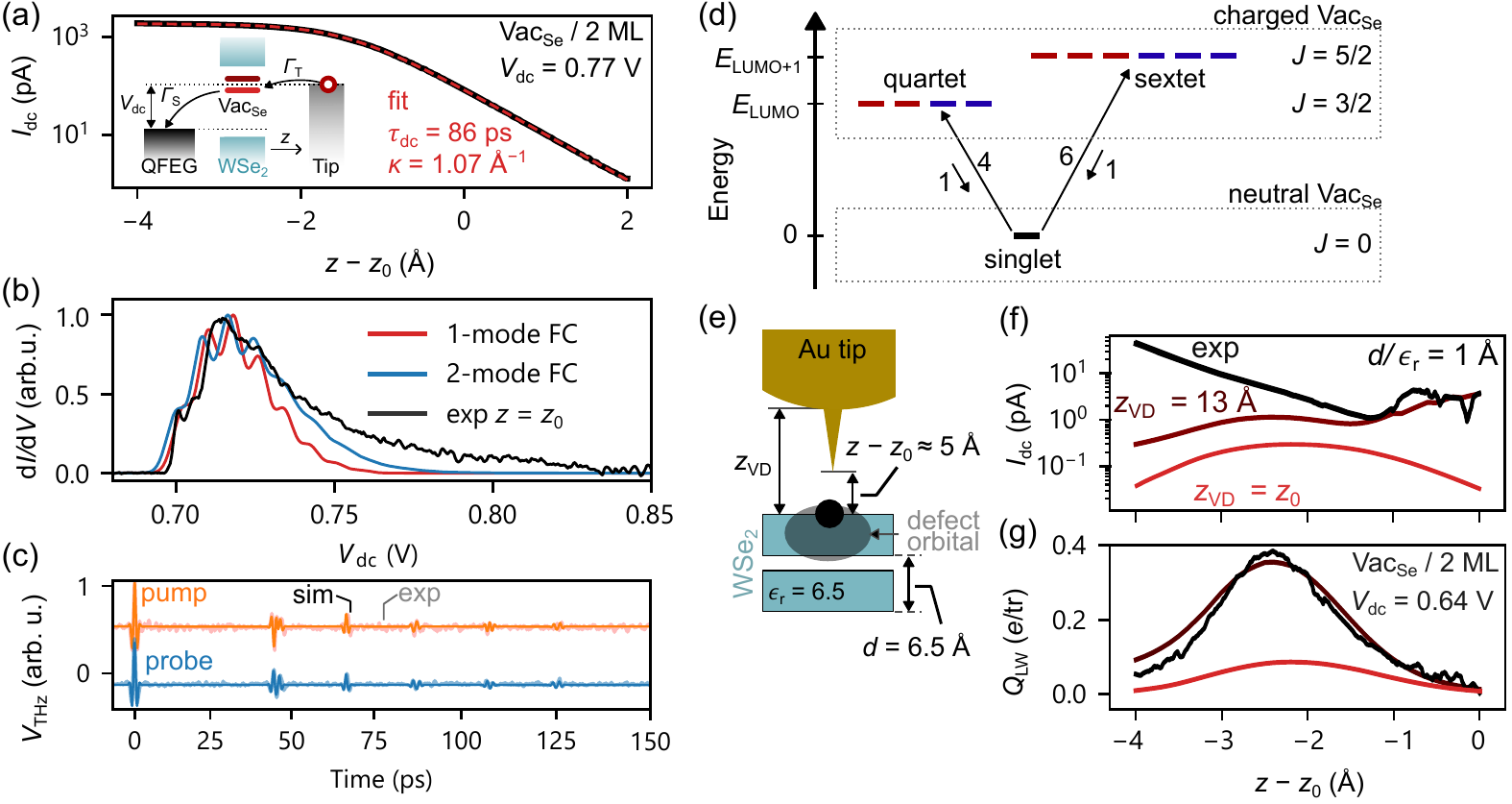}
\caption{\label{SI:fig:simulations}\textbf{Simulation of LW-rectified currents on \vacse{} in 2ML \wse.} (a)~The saturation of tunneling current for the \vacse{} LUMO resonance (\vdc$ = 0.77$\,V) reveals important experimental parameters, including the average charge-state lifetime $\tau_\text{dc}$, current at $z_0$, and the tunneling barrier $\kappa$, compare Fig. \ref{SI:fig:defect_characterization}f. (b)~Modeling of the LUMO resonances using a two-mode Franck--Condon model with a dominant Gaussian phonon modes at 8\,m$e$V spacing. The Franck--Condon parameters are $\hbar\omega_1=8$\,m$e$V, $S_1=2.2$, $n_1=8$, $\hbar\omega_2=20$\,m$e$V, $S_2=0.7$, $n_2=6$, where $\hbar\omega$ is the phonon mode's energy, $S$ is their Huang-Rhys factor, $n$ is the maximal amount of phonon excitations and a Gaussian line shape with 3\,m$e$V broadening is assumed. For simplicity, we assume the same vibronic broadening for the LUMO and LUMO+1 states. The d$I$/d$V$ measurement is from defect Fig.~\ref{SI:fig:defect_characterization}c. (c)~Analytic reconstruction of the THz near-field waveform including various reflections matching the experimentally measured pump and probe transient. (d)~Schematic illustration of multiplicity of \vacse{} states. High angular momentum reduces the probability of backward transitions from charged to neutral states. To reproduce the similar strength of LUMO and LUMO+1 an orbital-dependent coupling factor of 1/2 is required for the LUMO+1. (e)~A sketch of the parameters that influence the voltage drop, which are the dielectric constant $\epsilon_r$ of \wse, the graphene--defect distance $d$ and the defect--tip distance $z_\text{VD}$. The \vacse{} orbital extents throughout the whole \wse{} layer and thus we assume $d$\,=\,6.5\,\AA. The STM tip can be treated as a large sphere that acts as plate capacitor with an atomic protrusion that localizes the tunneling current. The distance between \wse{} and the protrusion is $z_0$ while the distance $z_\text{VD}$ influences the observed voltage drop. The voltage drop depends on $d/(\epsilon_r z_\text{VD})$, thus varying $d/\epsilon_r$ or $z_\text{VD}$ is equivalent.
(f)~The simulation from Fig.~\ref{fig4}d for $z_\text{VD}$\,=\,13\,\AA{} is compared to the case of $z_\text{VD}\approx z_0$. The significant influence of the choice of voltage drop is shown for $Q_\text{LW}(z)$ simulations. Experimental data in panels (a) and (b) was obtained on Vac$_\text{Se}^{2.4}$ and Vac$_\text{Se}^{2.2}$ in 2\,ML \wse{} , respectively (Extended Fig.~\ref{SI:fig:defect_characterization}).
 }
\end{figure}

\begin{itemize}
    \item Extended Figure~\ref{SI:fig:defect_characterization}: STM characterization of different \vacse{} with unique identifiers in the
name superscript
    \item Extended Figure~\ref{SI:fig:THz}: THz waveform and amplitude calibration on pristine 2\,ML \wse
    \item Extended Figure~\ref{SI:fig:FC}: Suppression of back tunneling via the Franck--Condon blockade
    \item Extended Figure~\ref{SI:fig:backtunneling}: THz amplitude spectroscopy and THz approach curves at \vacse{} in 1 and 2\,ML \wse
    \item Extended Figure~\ref{SI:fig:orbital}: Spatial dependence of the Coulomb blockade
    \item Extended Figure~\ref{SI:fig:simulations}: Simulation of LW-rectified currents on \vacse{} in 2\,ML \wse
\end{itemize}

\clearpage

\bibliographystyle{nature_bsc}
\bibliography{references}

\end{document}